\documentclass[12pt,a4paper]{article}
\pdfoutput=1
\usepackage{macros}
\usepackage{lipsum}

\newcommand\blfootnote[1]{%
\begingroup 
\renewcommand\thefootnote{}\footnote{#1}%
\addtocounter{footnote}{-1}%
\endgroup 
}

\newcommand{\YL}[1]{{\color{orange}{\textbf{YL:} #1}}}

\title{Quasinormal Modes of C-metric from SCFTs}

\author[1]{Yang Lei,}
\author[2,3,4]{Hongfei Shu,}
\author[5,6]{Kilar Zhang,}
\author[1]{Rui-Dong Zhu\blfootnote{*The authors are ordered purely alphabetically and should all be viewed as the co-first authors. }}
\affiliation[1]{Institute for Advanced Study \& School of Physical Science and Technology,\\ Soochow University, Suzhou 215006, China
}
\affiliation[2]{School of Physics and Microelectronics,
Zhengzhou University, Zhengzhou,\\ Henan 450001, China}
\affiliation[3]{Beijing Institute of Mathematical Sciences and Applications, Beijing 101408, China}
\affiliation[4]{Yau Mathematical Sciences Center, Tsinghua University, Beijing 100084, China}
\affiliation[5]{Department of Physics, Shanghai University, Shanghai 200444, China}
\affiliation[6]{Shanghai Key Lab for Astrophysics, Shanghai 200234, China}

\emailAdd{leiyang@suda.edu.cn}
\emailAdd{shuphy124@gmail.com}
\emailAdd{kilar@shu.edu.cn}
\emailAdd{rdzhu@suda.edu.cn}
\abstract{We study the quasinormal modes (QNM) of the charged C-metric, which physically stands for a charged accelerating black hole, with the help of Nekrasov's partition function of 4d $\cN=2$ superconformal field theories (SCFTs). 
The QNM in the charged C-metric are classified into three types: the photon-surface modes, the accelerating modes and the near-extremal modes, and it is curious how the single quantization condition proposed in \cite{Aminov:2020yma} can reproduce all the different families. 
We show that the connection formula encoded in terms of Nekrasov's partition function captures all these families of QNM numerically and recovers the asymptotic behavior of the accelerating and the near-extremal modes analytically. 
Using the connection formulae of different 4d $\cN=2$ SCFTs, one can solve both the radial and the angular parts of the scalar perturbation equation respectively. 
The same algorithm can be applied to the de Sitter (dS) black holes to calculate both the dS modes and the photon-sphere modes.}

\begin{document}
\Yboxdim5pt

\allowdisplaybreaks

\maketitle

\section{Introduction}

Ever since the discovery of the gravitational waves \cite{GW150914}, there has been a growing importance placed on conducting more precise studies to extract information from observations of gravitational waves from black holes and compact stars. These studies have gained even greater significance, especially in light of the rapid promotion of future space-based gravitational observation projects led by China \cite{Hu:2017mde,Ruan:2018tsw,TianQin:2015yph} and Europe \cite{LISA:2017pwj}. Among the various intriguing physical quantities, the quasinormal modes (QNM) \cite{vishveshwara1970scattering,Kokkotas:1999bd,Berti:2009kk,Konoplya:2011qq,Horowitz:1999jd} of black holes stand out. They play a crucial role in characterizing the ring-down phase of gravitational wave emissions during merger events and are potentially valuable for investigating various candidate solutions within the framework of general relativity and modified gravity theories.

The traditional studies of the QNM are mainly based on numerical approaches such as the higher-order WKB approximation \cite{Konoplya:2003ii}, isomonodromy method (See \cite{Cavalcante:2023rdy} for a recent review), Bender-Wu method \cite{Hatsuda:2019eoj,Eniceicu:2019npi} and the discretization of differential equation \cite{Gundlach:1993tp,Jansen:2017oag}. 
In a recent work \cite{Aminov:2020yma}, however, an interesting connection between the QNM and 4d $\cN=2$ supersymmetric gauge theories are found, and it is proposed that the B-cycle quantization condition that appears in the so-called Nekrasov-Shatashvili limit of the gauge theory  \cite{Nekrasov:2009rc} gives the QNM of the corresponding black hole, both satisfying the form of (confluent) Heun equations. Nekrasov's partition function, which determines the B-cycle quantization condition, has a closed-form analytic expression as an infinite sum over the number of instantons \cite{Nekrasov:2002qd, Nekrasov:2003rj}. This not only enables one to study the QNM numerically but may also allow one to extract more analytic information about the physical properties of the QNM. The discovery of this surprising connection between two completely different topics triggered a lot of intensive studies on the QNM and related physical observables using the gauge theory data \cite{Hatsuda:2020sbn,Hatsuda:2020iql,Bonelli:2021uvf,daCunha:2021jkm,Bianchi:2021mft,Nakajima:2021yfz,Hatsuda:2021gtn,Fioravanti:2021dce,Bonelli:2022ten,Bianchi:2022wku,Dodelson:2022yvn,Consoli:2022eey,Imaizumi:2022qbi,Fioravanti:2022bqf,Lisovyy:2022flm,Bhatta:2022wga,daCunha:2022ewy,Gregori:2022xks,Imaizumi:2022dgj,Bianchi:2022qph,Fioravanti:2023zgi,Cano:2023tmv,Bianchi:2023rlt,Fucito:2023plp,Giusto:2023awo,Aminov:2023jve,Kwon:2023ghu,Hatsuda:2023geo,Bhatta:2023qcl}. 

In this article, we apply the above method to a particular black hole geometry, known as the charged C-metric. 
It is an exact solution to the Einstein-Maxwell equation \cite{Weyl-cmetric}, and physically has an interpretation as an accelerating black hole \cite{Kinnersley-Walker}. 
The analytic extension of the charged C-metric is also interpreted as two separated black holes moving away from each other due to the cosmic string, while the cosmic string is characterized by a conical singularity.
There are several families of QNM in the charged C-metric \cite{Destounis:2020pjk}, including the photon-surface modes (PS modes), the accelerating modes and the near-extremal modes. 
It is then puzzling how the {\it merely one} quantum number appearing in the quantization condition proposed in \cite{Aminov:2020yma} can reproduce three families of QNM. To solve this puzzle, we further borrow the help of the Alday-Gaiotto-Tachikawa (AGT) duality \cite{Alday:2009aq}, which states a highly non-trivial correspondence between Nekrasov's partition function in gauge theories and the correlation function in 2d CFTs. In the context of 2d CFT, it is rather straightforward to derive the core ingredient of computing QNM, i.e. the connection formula \cite{Bonelli:2022ten} which bridges the solutions in the different boundaries. 
In the meanwhile, Nekrasov's partition function gives a machinery to compute the connection formula systematically. 
In this paper, we show that the apparent puzzle is solved by the connection formula approach, and we discuss about the potential problems that the unstable performance of B-cycle quantization conditions in capturing all the QNM.
As a similar phenomenon is also observed in dS black holes and their generalizations, we further apply the connection formula to the vector and tensor perturbations of the dS black hole before we conclude this article. 

This paper is organized as follows. 
In section \ref{s:qSW}, we briefly review how the QNM of black holes and 4d $\cN=2$ supersymmetric gauge theories in the Nekrasov-Shatashvili limit are related through the Heun type differential equation. 
The review also includes the necessary gauge theory tools used in this article, i.e. the quantization of the Seiberg-Witten curve proposed in \cite{Nekrasov:2009rc} and the AGT duality. 
In section \ref{s:c-metric}, we summarize the numerical results of the QNM in the charged C-metric based on \cite{Destounis:2020pjk} and the {\it QNMspectral} package \cite{Jansen:2017oag}. 
As there are three families of QNM in the charged C-metric, an apparent puzzle arises when one tries to apply the quantization condition proposed in \cite{Aminov:2020yma} to solve the QNM. In section \ref{s:connf}, we apply the connection formula approach to study the QNM in the charged C-metric both numerically and analytically, and provide strong evidence that the connection formula gives all the families of the QNM by solving both the radial and angular parts of the Klein-Gordon equation. The relation between the B-cycle quantization condition and the connection formula will also be discussed and we further apply the same method to the dS black hole which also contains the photon-sphere mode and the dS modes.

\section{Quantized Seiberg-Witten approach to QNM}\label{s:qSW}

In this section, we provide a brief review on the interesting connection found in \cite{Aminov:2020yma} between the (quantized) Seiberg-Witten (SW) theory of 4d $\cN=2$ supersymmetric gauge theories and the quasinormal modes (QNM) of black holes, and collect necessary tools for later usage in this article. 

\subsection{Quasinormal modes of black holes}

There are a few salient features of
Quasinormal modes (QNM).
First of all, they are different from normal modes as QNM are damped exponentially.
This feature means the QNM of black holes are complex and contain negative imaginary part, which is due to the presence of event horizon. 
This negative nature of imaginary part of QNM has been confirmed by the explicit computations of QNM of Schwarzchild black holes \cite{vishveshwara1970scattering}, Kerr black holes \cite{press1973perturbations} and Reissner-Nordstr{\"o}m black holes \cite{moncrief1974stability,Cavalcante:2021scq}.
The computations of QNM were also done in spacetimes with more fruitful structures, such as the ones with cosmological constant \cite{Horowitz:1999jd,BarraganAmado:2018zpa,Amado:2020zsr,Amado:2021erf,Novaes:2018fry,Tattersall:2018axd,Cardoso:2003sw,Cardoso:2003cj,Cardoso:2017soq}, and also the C-metric \cite{Destounis:2020pjk} charaterizing the accelerating in the geometry.

The second feature is that the QNM are independent of initial data which triggers the black hole perturbation. 
This nature makes the QNM as the characteristic feature of black hole.
Just as the hydrogen atom is characterized by its spectrum, the QNM is also the fingerprint spectrum to characterize the black hole data, including the masses, angular momenta and charges.

The perturbation of black hole is described by the field equation in the classical background. 
For example, the scalar perturbation of the black hole is described by 
Klein Gordon equation. 
Similarly, perturbations with other spin$-\frac{1}{2}$ or spin$-1$ or higher \cite{Cavalcante:2021scq,Lin:2014ata,Amado:2020zsr} are also studied in various background. 
The results are applied to compute the absorption rate.

The QNM are defined to be the solutions with ingoing boundary condition at the black hole horizon and outgoing modes near the infinity.
In many exact black hole backgrounds, such as Schwarzchild, Kerr, and Reissner-Norstrom etc, the partial differential equations describing the black hole perturbations
are separable \cite{Kubiznak:2008qp, Frolov:2017kze,Wu:2009ug,Castro:2010fd}. (i.e. can be reduced to ordinary differential equations in radial part and angular part respectively).
The equations in radial direction are generically Heun type differential equations or their confluent reduced types \cite{Aminov:2020yma} :
\begin{equation} 
    \frac{{\rm d}^2}{{\rm d}z^2}\tilde{\psi}(z)+\lt[\frac{\tilde{\gamma}}{z}+\frac{\tilde{\delta}}{z-1}+\frac{\tilde{\epsilon}}{z-t}\rt]\frac{{\rm d}}{{\rm d}z}\tilde{\psi}(z)+\frac{\tilde{\alpha} \tilde{\beta} z-\tilde{q}}{z(z-1)(z-t)}\tilde{\psi}(z)=0,\label{eq-Heun}
\end{equation}
with $\tilde{\epsilon}=1 +\tilde{\alpha}+\tilde{\beta}-\tilde{\gamma}-\tilde{\delta}$.  
The problems of computing QNM are then reduced to study the eigenvalue problem of this differential equation.

The Heun equation is at the center of our discussion. 
It appears in physical topics from many different areas, including quantization of Seiberg-Witten curve, and BPZ equations in conformal field theory. 
Thus the algorithm developed for solving this equation works as a nexus to connect different problems.

\subsection{Seiberg Witten curve and its quantization}

A well-known construction of $\cN=2$ gauge theories, called the class-$\cS$ construction, was proposed in \cite{Gaiotto:2009we} by compactfying 6d $\cN=(2,0)$ theory on a punctured Riemann surface $\Sigma$. In this picture, the Seiberg-Witten curve, which determines the low-energy effective action of the supersymmetric gauge theory \cite{Seiberg:1994rs,Seiberg:1994aj}, is given by a ramified double cover of $\Sigma$, and is of course characterized by its genus and the type of singularities defined on it. The Seiberg-Witten curve for SU(2) theory with $N_f=N_{f_1}+N_{f_2}$ flavors is known to take the form, 
\begin{equation}
   \lt(\frac{P_1(x)}{y}+\bar{\Lambda}P_2(x)y\rt)=x^2-u, \label{SW-curve}
\end{equation}
where $\bar{\Lambda}$ is a function of the instanton counting parameter $\mathfrak{q}=e^{2\pi i\tau}$ (with $\tau=\frac{4\pi i}{g^2}+\frac{\theta}{2\pi}$ as the complexified coupling combining the gauge coupling $g$ and the theta angle $\theta$) depending on $N_f$,  and 
\begin{align}
    P_1(x)=\prod_{i=1}^{N_{f_1}}(x-m_i),\quad P_2(x)=\prod_{i=N_{f_1}+1}^{N_{f_1}+N_{f_2}}(x-m_i),
\end{align}
are polynomials that encode the mass information of fundamental and anti-fundamental hypermultiplets. 
In particular in the case of $N_f=4$, which we mainly consider in this article, we simply have $\bar{\Lambda}=\mathfrak{q}$. 
The case with $N_f=3,2$ are considered in \cite{Aminov:2020yma}. 
When we switch on an $\Omega$-background parameter (i.e. the Nekrasov-Shatashvili limit, or the NS limit), say $\epsilon_1=\hbar$, it is still possible to solve the gauge theory by quantizing the Seiberg-Witten theory \cite{Nekrasov:2009rc} to a differential equation. The quantization condition $\lt[x,\log y\rt]=\hbar$ can be realized by replacing $x=\hbar y\partial_y$ and $P(x)y=yP(x+\hbar)$ in a symmetric quantization manner, 
\begin{equation}
    \lt(\bar{\Lambda}y^{\frac{1}{2}}P_2\lt(x\rt)y^{\frac{1}{2}}+P_0(x)+y^{-\frac{1}{2}}P_1\lt(x\rt)y^{-\frac{1}{2}}\rt)\Psi(y)=0.
\end{equation}
The quantized version of curve \eqref{SW-curve} then becomes \cite{Tai:2010ps,Maruyoshi:2010iu,Zenkevich:2011zx}
\begin{equation}
    \lt(\bar{\Lambda}y^2P_2\lt(x+\frac{\hbar}{2}\rt)+yP_0(x)+P_1\lt(x-\frac{\hbar}{2}\rt)\rt)\Psi(y)=0,\label{eq:qSW}
\end{equation}
where $P_0(x):=x^2+\bar{\Lambda}\delta_{N_f,3}-\hat{u}+\bar{\Lambda}p_0(x)\delta_{N_f,4}$, and 
\begin{equation}
    p_0(x)=x^2-\lt(x+\frac{\hbar}{2}\rt)\sum_im_i+\hat{u}+\sum_{i<j}m_im_j+\frac{\hbar^2}{2}.
\end{equation}
In this article, we focus on the case $N_{f_1}=N_{f_2}=2$ (i.e. $N_f=4$), and in this case $\bar{\Lambda}=\mathfrak{q}$ is simply given by the instanton counting parameter. One can bring the above differential equation for $N_f=4$ into the standard form of a Schr\"odinger equation, 
\begin{equation}\label{eq:SW-eq}
 \lt(\hbar^2\frac{{\rm d}^2}{{\rm d}z^2}+Q_{\rm SW}(z)\rt)\psi(z)=0,
\end{equation}
with $z=-y^{-1}$, $t=\mathfrak{q}$, 
\begin{equation}
    Q_{\rm SW}(z)=\frac{\frac{1}{4}-a_{0}^{2}}{z^{2}}+\frac{\frac{1}{4}-a_{1}^{2}}{(z-1)^{2}}+\frac{\frac{1}{4}-a_{t}^{2}}{(z-t)^{2}}-\frac{\frac{1}{2}-a_{0}^{2}-a_{1}^{2}-a_{t}^{2}+a_{\infty}^{2}+u}{z(z-1)}+\frac{u}{z(z-t)},\label{QSW}
\end{equation}
where the coefficients are determined by
\begin{align}
    &a^2_\infty=\frac{1-\hbar^2+(m_3-m_4)^2}{4},\\
    &a_0^2=\frac{1-\hbar^2+(m_1-m_2)^2}{4},\\
    &a_1^2=\frac{1-\hbar^2+(m_3+m_4+2\hbar)^2}{4},\\
    &a_t^2=\frac{1-\hbar^2+(m_1+m_2-2\hbar)^2}{4},\\
    &u=-\hat{u}+\frac{m_1^2+m_2^2}{2}+\frac{(m_1+m_2)(m_3+m_4)\mathfrak{q}}{2(1-\mathfrak{q})}\cr
    &+(m_1+m_2)\hbar+\frac{\mathfrak{q}\hbar}{2}\sum_{i=1}^4m_i+\frac{3-\mathfrak{q}}{4(1-\mathfrak{q})}\hbar^2.
\end{align}
The function $\psi(z)$ is related to $\Psi$ by eliminating the first order derivative term in the differential equation: 
\begin{align}\label{eq:transforamtion-psi-Psi}
    \psi(z)=\exp\lt(\frac{1}{\hbar}\int \frac{\mathfrak{q}(m_1+m_2)(z-1)-(m_3+m_4)z(z-\mathfrak{q})-(z+(z-2)\mathfrak{q})\hbar}{z(z-1)(z-\mathfrak{q})}{\rm d}z\rt)\cr
    \times \Psi(-z^{-1}).
\end{align} 
Interestingly the resulting equation \eqref{eq:SW-eq} under the transformation \eqref{eq:transforamtion-psi-Psi} (with $\hbar=1$) is precisely the famous Heun equation \eqref{eq-Heun}.
To see this explicitly, one should make following identification between parameters
\begin{align}
    a_0^2=\frac{(1-\tilde{\gamma})^2}{4},\quad  a_1^2=\frac{(1-\tilde{\delta})^2}{4},\quad a_t^2=\frac{(\tilde{\alpha}+\tilde{\beta}-\tilde{\gamma}-\tilde{\delta})^2}{4},\quad a_\infty^2=\frac{(\tilde{\alpha}-\tilde{\beta})^2}{4},
\end{align}
\begin{equation}
    u=\frac{\tilde{\gamma}\tilde{\epsilon}-2\tilde{q}+2\tilde{\alpha}\tilde{\beta} t-t\tilde{\epsilon}(\tilde{\gamma}+\tilde{\delta})}{2(t-1)},
\end{equation}
and the identification between functions 
\begin{equation}
    \tilde{\psi}(z)=\exp\lt(-\int\frac{z^2(1+\tilde{\alpha}+\tilde{\beta})-z(1+\tilde{\alpha}+\tilde{\beta}-\tilde{\delta}+(\tilde{\gamma}+\tilde{\delta})t)+t\tilde{\gamma}}{2z(z-1)(z-t)}{\rm d}z\rt)\psi(z).\label{conv-eq}
\end{equation}

The pole-structure appeared in \eqref{QSW} is exactly what one expects from the class-$\cS$ construction, i.e. the Riemann surface $\Sigma$ has four regular punctures for 4d SU(2) theory with $N_f=4$ \cite{Gaiotto:2009we}. It is then natural to construct the solutions to the Heun equation  in terms of the instanton partition function (in the NS limit).
This is exactly the idea presented in \cite{Aminov:2020yma} to solve the quasinormal modes of black holes with the instanton partition function.

\subsection{AGT duality to conformal blocks}

The explicit expression of the solutions written in terms of the instanton partition function can be found via yet another connection with the Heun equation, the Alday-Gaiotto-Tachikawa (AGT) relation \cite{Alday:2009aq}. 
It states an intriguing duality between 4d $\cN=2$ SU(2) gauge theories and 2d Liouville CFTs. More precisely, the instanton partition function in the gauge theory with $N_f=4$ flavors is mapped to the 4-pt conformal block with conformal weights $\{\Delta_i\}_{i=1}^4$  of the external lines and $\Delta$ of the internal line. The conformal block is usually schematically represented as follows, 
\begin{align}
     \begin{tikzpicture}
        \draw[ultra thick] (-1,1)--(0,0);
        \draw[ultra thick] (-1,-1)--(0,0);
        \draw[ultra thick] (2,0)--(0,0);
        \draw[ultra thick] (2,0)--(3,1);
        \draw[ultra thick] (2,0)--(3,-1);
        \node at (-1,1.2) [left,circle,draw=blue!50,fill=blue!30] {$\Delta_1^{(0)}$};
        \node at (-1,-1) [left,circle,draw=blue!50,fill=blue!30] {$\Delta_2^{(t=\mathfrak{q})}$};
        \node at (3,1.2) [right,circle,draw=blue!50,fill=blue!30] {$\Delta_3^{(1)}$};
        \node at (3,-1) [right,circle,draw=blue!50,fill=blue!30] {$\Delta_4^{(\infty)}$};
        \node at (1,0) [above] {$\Delta$};
    \end{tikzpicture}
\end{align}
Here we introduce the convention $\Delta^{(x_i)}_i$ to stand for operators inserted at spacetime point $x_i$ with conformal dimension $\Delta_i$.
The dictionary between conformal blocks $\mathfrak{F}$ and the data of supersymmetric gauge theory on the $\Omega$-background with two parameters $\epsilon_{1,2}$ is given by 
\begin{align}
\begin{split}
& c=1+6\cQ^2,\quad \cQ=b+b^{-1},\quad \epsilon_1=b,\quad \epsilon_2=b^{-1},\\
& \Delta_i=\frac{\cQ^2}{4}-\alpha_i^2,\quad \Delta=\frac{\cQ^2}{4}-\alpha^2,\quad \bar{a}:=b\a=b a,\\
& m_1=\alpha_1+\alpha_2+\frac{\cQ}{2},\quad m_2=-\alpha_1+\alpha_2+\frac{\cQ}{2},\\
& m_3=\alpha_3+\alpha_4+\frac{\cQ}{2},\quad m_4=\alpha_3-\alpha_4+\frac{\cQ}{2}.
 \end{split}
\end{align}
The statement of the AGT relation can thus be 
expressed as\footnote{The conformal block defined here differs from the one used in the usual context of Virasoro algebra by an overall weight factor $z^{\Delta-\Delta_2-\Delta_1}$. In particular, in the case that corresponds to the hypergeometric function with $\Delta=\Delta_{0\theta}$, $\Delta_1=\Delta_0$, $\Delta_2=\Delta_{2,1}$, the overall factor is given by $z^{\frac{bQ}{2}+\theta b\alpha_0}$.} 
\begin{align}
 &   \mathfrak{F}(\Delta_4,\Delta_3,\Delta,\Delta_2,\Delta_1;\mathfrak{q})=\mathfrak{q}^{\Delta-\Delta_2-\Delta_1}(1-\mathfrak{q})^{-2(\a_2+\frac{\cQ}{2})(\frac{\cQ}{2}-\a_3)}\cr
    &\times Z^{inst}_{SU(2)\ N_f=4}(a,\{2\epsilon_+-m_1,2\epsilon_+-m_2,m_3,m_4\};\mathfrak{q}),\label{eq:AGT}
\end{align}
where a U(1) factor is suppressed from the instanton partition function \cite{Alday:2009aq}. Note that the instanton counting parameter $\mathfrak{q}$ is identified as one of the positions $t$ where one primary field with conformal weight $\Delta_2$ is inserted. 

The 4-pt (holomorphic) correlation function with a degenerate field $\Phi(z)$ of conformal weight $\Delta_{2,1}=-\frac{1}{2}-\frac{3b^2}{4}$ inserted satisfies the so-called BPZ equation: 
\begin{align}
    \lt(b^{-2}\partial_z^2+\frac{\Delta_2}{(z-1)^2}-\frac{\sum_{i=1}^3\Delta_i+\Delta_{2,1}+t\partial_t+z\partial_z-\Delta_4}{z(z-1)}+\frac{\Delta_3}{(z-t)^2}\rt.\cr
    \lt.+\frac{t\partial_t}{z(z-t)}-\frac{\partial_z}{z}+\frac{\Delta_1}{z^2}\rt)\bra{\Delta_4}V_2(1)V_3(t)\Phi(z)\ket{\Delta_1}=0.\label{eq:4pt-BPZ}
\end{align}
The semiclassical limit is defined by taking $b\rightarrow 0$ with $b\alpha_i$ finite to satisfy
\begin{equation}
b \alpha_1 \to a_0, \quad b\alpha_2 \to a_1, \quad b \alpha_3 \to a_t, \quad b \alpha_4 \to a_\infty
\end{equation}
The BPZ equation in this limit reduces to the Schr\"odinger form similar to Heun equation \eqref{eq:SW-eq} for $\hbar=1$.
The correlation functions are known to be decomposed into conformal blocks as
\begin{align}
\begin{split}
& \bra{\Delta_4}V_3(1)V_2(t)\Phi(z)\ket{\Delta_1}=\sum_{\theta=\pm}\int{\rm d}\alpha C^{\alpha_{0\theta}}_{\alpha_{2,1}\alpha_0}C^{\alpha}_{\alpha_t\alpha_{0\theta}}C_{\alpha_\infty\alpha_1\alpha}\\
    &\times\lt|\mathfrak{F}(\Delta_4,\Delta_3,\Delta,\Delta_2,\Delta_{1\theta},\Delta_{2,1},\Delta_1;t,z/t)\rt|^2,\\
 &   \bra{\Delta_4}V_3(1)V_2(t)\ket{\Delta_1}=\sum_{\theta=\pm}\int{\rm d}\alpha C^{\alpha}_{\alpha_t\alpha_0}C_{\alpha_\infty\alpha_1\alpha}\lt|\mathfrak{F}(\Delta_4,\Delta_3,\Delta,\Delta_2,\Delta_1;t)\rt|^2.
\end{split}
\end{align}
The conformal block is divergent in this semiclassical limit, namely
\begin{equation}
    \mathfrak{F}(\Delta_4,\Delta_3,\Delta,\Delta_2,\Delta_1;t)=t^{\Delta-\Delta_2-\Delta_1}e^{b^{-2}F(t)+\cO(1)},
\end{equation}
and the $z$-dependence in the 4-pt conformal block with degenerate insertion (we call it 5-pt degenerate conformal block), 
\begin{align}
    \begin{tikzpicture}
        \draw[ultra thick] (-1,1)--(0,0);
        \draw[ultra thick,dashed] (-1,-1)--(0,0);
        \draw[ultra thick] (0,-1)--(1,0);
        \draw[ultra thick] (2,0)--(0,0);
        \draw[ultra thick] (2,0)--(3,1);
        \draw[ultra thick] (2,0)--(3,-1);
        \node at (-1,1) [left,circle,draw=blue!50,fill=blue!30] {$\Delta_1^{(0)}$};
        \node at (-1,-1) [left,circle,draw=red!50,fill=red!30] {$\Delta_{2,1}^{(z)}$};
        \node at (0,-1) [below,circle,draw=blue!50,fill=blue!30] {$\Delta_{2}^{(t)}$};
        \node at (3,1) [right,circle,draw=blue!50,fill=blue!30] {$\Delta_3^{(1)}$};
        \node at (3,-1) [right,circle,draw=blue!50,fill=blue!30] {$\Delta_4^{(\infty)}$};
        \node at (1.5,0) [above] {$\Delta$};
        \node at (0.5,0) [above] {$\Delta_{1,\pm}$};
        \node at (7.5,0) {$=\mathfrak{F}(\Delta_4,\Delta_3,\Delta,\Delta_2,\Delta_{1,\pm},\Delta_{2,1},\Delta_1;t,z/t)$};
    \end{tikzpicture}
\end{align}
where $\Delta_{1,\pm}=\frac{\cQ^2}{4}-\lt(\alpha_1\mp \frac{b}{2}\rt)^2$, turns out to be subleading, 
\begin{equation}
    \mathfrak{F}(\Delta_4,\Delta_2,\Delta,\Delta_3,\Delta_{1,\pm},\Delta_{2,1},\Delta_1;t,z/t)=t^{\Delta-\Delta_3-\Delta_{1,\pm}}z^{\frac{b\cQ}{2}\pm b\a_1}e^{b^{-2}F(t)+W(z/t,t)+\cO(b^2)}.\label{leading-confb}
\end{equation}
One way to cure this divergence is to follow \cite{Bonelli:2022ten} to regularize the 5-pt degenerate conformal block with the 4-pt conformal block, 
\begin{align}
    \cF_\pm(z;t):=\lim_{b\rightarrow 0}\frac{\mathfrak{F}(\Delta_4,\Delta_2,\Delta,\Delta_3,\Delta_{1,\pm},\Delta_{2,1},\Delta_1;t,z/t)}{\mathfrak{F}(\Delta_4,\Delta_2,\Delta,\Delta_3,\Delta_1;t)}\cr
    =t^{\mp a_1}z^{\frac{1}{2}\pm a_1}e^{\mp\frac{1}{2}\partial_{a_1}F(t)}\lt(1+\cO(t,z/t)\rt).\label{semi-reg}
\end{align}
During the regularization process, one of the conformal blocks in the correlator is chosen out by specializing the internal channel $\Delta$. 
$\cF_\pm$ give two solutions to the Heun equation \eqref{eq:SW-eq}, with $\hbar=1$, and it is easy to check that 
\begin{equation}
    u=\lim_{b\rightarrow 0}b^2t\partial_t\log\mathfrak{F}(\Delta_4,\Delta_2,\Delta,\Delta_3,\Delta_1;t),\label{u-cond}
\end{equation}
in this Heun equation can be computed only from the 4-pt conformal block or the instanton partition function of 4d SU(2) gauge theory with $N_f=4$.

This argument can be extended to the 6-pt degenerate conformal block, which turns out to be useful in the calculation of the angular part of the charged C-metric considered in this article. In Appendix \ref{a:sol}, we give the explicit expressions of Nekrasov's partition functions and the conformal blocks, which will later be used in the concrete calculations of the QNM.

\subsection{WKB approach and quantization conditions}\label{s:WKB}

Given a second-order differential equation, one can solve it in the expansion of the Planck constant $\hbar$ with the Wenzel-Kramers-Brillouin (WKB) method. 
The solution to the quantized SW curve \eqref{eq:qSW} can be found as  
\begin{equation}
	\lt(H(z,p)-E\rt)\exp\lt(\frac{i}{\hbar}\int^z\hat{\lambda}(z',\hbar,E)\rt)=0,
\end{equation}
with the identification $z=\log y$ and $p=\hbar\partial_z=\hbar y\partial_y=x$. 
The eigenenergy $E$ can also be identified as a function of the parameter $\hat{u}$. 
The differential one-form $\hat{\lambda}(z,\hbar,E)$ can be solved recursively as an infinite series of $\hbar$, 
\begin{equation}
	\hat{\lambda}(y,\hbar,E)=\sum_{k=0}^\infty \hat{\lambda}_k\hbar^k,
\end{equation}
and the leading order contribution $\hat{\lambda}_0$ in the equation \eqref{eq:qSW} is given by the value of $p=x$ (at the classical level) in the Seiberg-Witten curve \eqref{SW-curve},
\begin{equation}
    \hat{\lambda}_0=\frac{x}{y}{\rm d}y.
\end{equation}
It is nothing but the celebrated Seiberg-Witten differential equipped on the SW curve. There are two branch cuts in the Seiberg-Witten curve (denoted as $C_{i=1,2}$) and the curve is topologically equivalent to a torus. In the Seiberg-Witten theory, the integrals around $C_i$ are related to the Coulomb branch parameters $\mathfrak{a}_i$, 
\begin{equation}
    \oint_{C_i}\hat{\lambda}_0=\mathfrak{a}_i,\quad i=1,2.
\end{equation}
Imposing the SU(2)-condition on the theory restricts the parameters to $\mathfrak{a}_1=-\mathfrak{a}_2=:\mathfrak{a}$. 
There are one A-cycle and one B-cycle on a torus, so it is usually identified as one of the cycles around the branch cut, say $C_1$, and a cycle connecting two endpoints of different branch cuts. 
The A-cycle and B-cycle integrals are related to the gauge-theory data via \cite{Seiberg:1994rs,Seiberg:1994aj} (A good textbook on the Seiberg-Witten curve can be found in \cite{Tachikawa:2013kta}) 
\begin{align}
    \oint_A\hat{\lambda}_0=\mathfrak{a},\quad \oint_B\hat{\lambda}_0=\partial_{\mathfrak{a}}{\cal F}_0,
\end{align}
where ${\cal F}_0$ is the prepotential of the gauge theory and can be found in the classical limit $\epsilon_{1,2}\rightarrow 0$ from the Nekrasov partition function \cite{Nekrasov:2002qd,Nekrasov:2003rj}, 
\begin{equation}
    {\cal F}_0=\lim_{\epsilon_{1,2}\rightarrow 0}\epsilon_1\epsilon_2\log Z_{\rm SU(2)}.
\end{equation}
Turning on one of the $\Omega$-background parameter $\epsilon_1=\hbar$ introduces quantum corrections to the gauge theory, and correspondingly in the quantum mechanics approach, we shall use $\hat{\lambda}:=\lambda(y,\hbar,E){\rm d}z$ as the quantum-corrected differential one-form to generate the quantum version of the A-cycle and B-cycle integrals, which are usually referred to as the quantum A-period and the quantum B-period. 
\begin{align}
    \Pi^{\rm WKB}_A=\oint_{A}\hat{\lambda}=\sum_{k=0}^\infty \hbar^k\oint_A\lambda_k{\rm d}z,\\
    \Pi^{\rm WKB}_B=\oint_{B}\hat{\lambda}=\sum_{k=0}^\infty \hbar^k\oint_B\lambda_k{\rm d}z. 
\end{align}
In \cite{Nekrasov:2009rc}, it is shown that the Nekrasov partition function in the NS limit ($\epsilon_1=\hbar$, $\epsilon_2\rightarrow 0$) gives  
\begin{equation}
    \Pi^{\rm WKB}_B=\partial_a {\cal F}^{\rm NS},\quad {\cal F}^{\rm NS}=-\hbar\lim_{\epsilon_2\rightarrow 0}\epsilon_2\log Z_{\rm SU(2)},
\end{equation}
and the quantization condition imposed on the quantum B-period gives rise to the Bethe ansatz equation in the corresponding quantum integrable model.

\begin{figure}
    \centering
    \includegraphics[width=8cm]{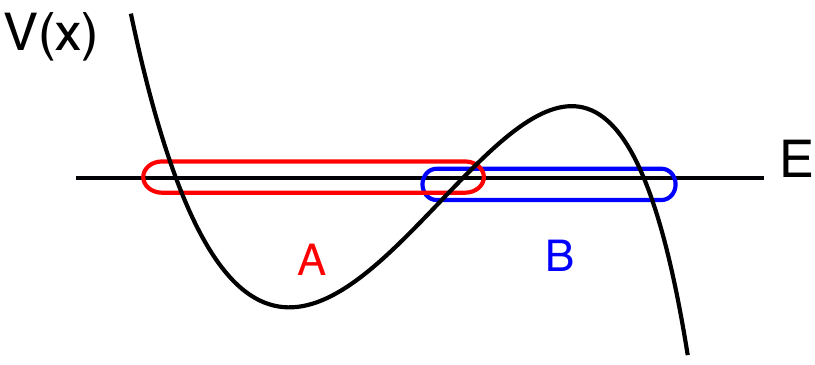}
    \caption{The A-cycle and B-cycle as classically allowed and forbidden regions in quantum mechanics. }
    \label{fig:AB-cycle}
\end{figure}

In the WKB picture, however, the quantum periods are equivalent to (twice) the integrals between turning points, and thus imposing the bound-state boundary condition or the resonance boundary condition respectively correspond to the quantization of the quantum A-period and the B-period (see Figure \ref{fig:AB-cycle}). As will be described in more details later, the QNM satisfy an ingoing boundary condition at the event horizon and an outgoing boundary condition at infinity or some horizon such as the cosmological horizon (in dS space) far away from the black hole. 
It can thus be interpreted as resonant states in an inverted-harmonic-oscillator-like potential, and one naturally expects the quantization of the B-cycle integral in the gauge theory gives the QNM as argued in \cite{Aminov:2020yma}.
For a comprehensive review, see \cite{Hatsuda:2021gtn}. 

\section{QNM in the charged C-metric}\label{s:c-metric}

In this article, we mainly focus on a special example, known as the charged C-metric, and try to compute its QNM from the corresponding quantized Seiberg-Witten differential equation. 
The charged C-metric was first discovered in \cite{Weyl-cmetric} as an exact solution to the Einstein-Maxwell equation, and later in \cite{Kinnersley-Walker}, it was interpreted physically as two causally separated black holes accelerating in opposite directions due to the influence of a cosmic string. 
The presence of acceleration in AdS spacetime (referred to AdS C-metric) also has novel interesting effect in black hole thermodynamics and holography. 
These were discussed in four dimensional AdS C-metric models \cite{Anabalon:2018ydc,Boido:2022iye,Cassani:2021dwa} and also three dimensional accelerating AdS spacetimes \cite{Arenas-Henriquez:2022www,Arenas-Henriquez:2023hur}.

In spherical-type coordinate the charged C-metric is more convenient for taking zero accelerating limit, which leads to Schwarzchild-like black holes.
The metric covering one charged accelerating black hole can be written as \cite{Griffiths:2005qp,griffiths2009,Griffiths:2006tk}
\begin{equation}
    {\rm d}s^2=\frac{1}{(1-\alpha r\cos\theta)^2}\lt(-f(r){\rm d}t^2+\frac{{\rm d}r^2}{f(r)}+\frac{r^2{\rm d}\theta^2}{P(\theta)}+P(\theta)r^2\sin^2\theta{\rm d}\varphi^2\rt),
\end{equation}
where 
 \begin{align}
    f(r)=\lt(1-\frac{2M}{r}+\frac{Q^2}{r^2}\rt)(1-\alpha^2r^2),\\
    P(\theta)=1-2\alpha M\cos\theta+\alpha^2Q^2\cos^2\theta,
\end{align}
with $M$, $Q$ and $\alpha$ respectively the mass, the charge and the accelerating parameter of the black hole. 

The scalar perturbation of the charged C-metric obeys the Klein-Gordon equation given by 
\begin{align}
    -f(r)^{-1}r^2\partial_t^2\tilde{\psi}+\partial_r(r^2f(r)\partial_r\tilde{\psi})+\frac{1}{\sin\theta}\partial_\theta(P(\theta)\sin\theta\partial_\theta\tilde{\psi})+\frac{\partial^2_\varphi\tilde{\psi}}{\sin^2\theta P(\theta)}\cr
    +\frac{1}{6}(r^2f^{\prime\prime}(r)+4rf'(r)+2f(r)+P^{\prime\prime}(\theta)+3\cot\theta P'(\theta)-2P(\theta))\tilde{\psi}=0.
\end{align}
By separation of variables of $\tilde{\psi}$ as  
\begin{equation}
    \tilde{\psi} (t,r,\theta,\varphi) = e^{-i\omega t}e^{im\varphi}\frac{\phi(r)}{r}\chi(\theta),
\end{equation}
one finds the radial equation and the angular equation respectively given by  
\begin{align}
    \frac{{\rm d}^2\phi(r)}{{\rm d}r^2_\ast}+(\omega^2-V_r)\phi(r)=0, \label{eq:BHODE-r}\\
    \frac{{\rm d}^2\chi(\theta)}{{\rm d}z^2}-(m^2-V_\theta)\chi(\theta)=0,\label{eq:BHODE-theta}
\end{align}
where 
\begin{equation}
    {\rm d}r_\ast=\frac{{\rm d}r}{f(r)},\quad {\rm d}z=\frac{{\rm d}\theta}{P(\theta)\sin\theta},\quad m=m_0P(\pi),\quad m_0\in\mathbb{Z},
\end{equation}
and 
\begin{align}
    V_r&=f(r)\lt(\frac{\lambda}{r^2}-\frac{f(r)}{3r^2}+\frac{f'(r)}{3r}-\frac{f^{\prime\prime}}{6}\rt),\\
    V_\theta&=P(\theta)\lt(\lambda\sin^2\theta-\frac{P(\theta)\sin^2\theta}{3}+\frac{\sin\theta\cos\theta P'(\theta)}{2}+\frac{\sin^2\theta P^{\prime\prime}(\theta)}{6}\rt).
\end{align}
We notice that the angular part in the limit $\alpha\rightarrow 0$ reduces to the equation for the spherical harmonics, and we have $\lambda\rightarrow \ell(\ell+1)+1/3$ for $^\exists \ell\in\mathbb{N}$ ($m_0$ corresponds to the $z$-component of the angular momentum in this limit). For more general $\alpha$, it gets highly non-trivial corrections whose analytic form is yet unknown at the moment. 
There are various kinds of different numerical methods to compute $\lambda$ and its value at different parameters are listed in Appendix \ref{a:angular}. 
Moreover the angular equation \eqref{eq:BHODE-theta} can be mapped to a Schr\"odinger equation with five regular singularities and thus can be solved from an SU(2)$\times$SU(2) quiver gauge theory. Some details are also provided in Appendix \ref{a:angular} and more will be reported in our follow-up paper.

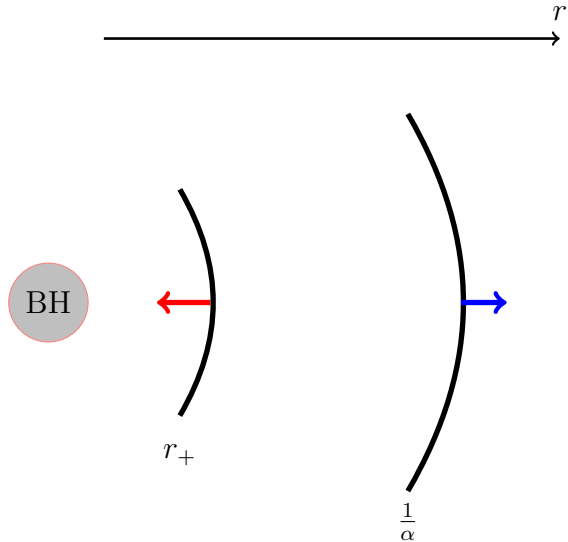
\begin{figure}
\centering
\begin{tikzpicture}
\draw [line width =2pt] (1,0)  to [out =300,in =60] (1,-3);
\draw [line width =2pt] (4,1)  to [out =300,in =60] (4,-4);
\draw [line width =2pt,->,red] (1.4,-1.5)  to  (0.7,-1.5);
\draw [line width =2pt,->,blue] (4.7,-1.5)  to  (5.3,-1.5);
	\node [above ] at (1,-3.8) {$r_+$};
	\node [above ] at (6,2.1) {$r$};
		\node [above ] at (4,-4.8) {$\frac{1}{\alpha}$};
\draw [line width =1 pt,->] (0,2)  to  (6,2);
 \node at (-0.2,-1.5) [left,circle,draw=red!50,fill=gray!50!white] {BH};
\end{tikzpicture}
\caption{The boundary condition of C-metric geometry.}
\label{fig:boundarycondition}
\end{figure}

The QNM are defined as special values of the frequencies $\omega$ in the radial equation satisfying the following boundary conditions, 
\begin{equation}\label{eq:bdy-BH}
    \phi(r)\sim\begin{cases}
e^{-i\omega r_{\ast}} & r_{\ast}\to-\infty,\quad r\to r_{+}\\
e^{i\omega r_{\ast}}& r_{\ast}\to+\infty,\quad r\to \frac{1}{\alpha}
\end{cases}.
\end{equation}
As $r_\ast$ is often called the tortoise coordinate and physically corresponds to the coordinate of a static observer outside the outer horizon, the solution of the form $e^{-i\omega t-i\omega r_\ast}$ describes a plane wave moving towards the negative $r_\ast$-direction, i.e. into the outer horizon direction, and $e^{-i\omega t+i\omega r_\ast}$ describes a plane waving moving towards the infinity, i.e. out to the acceleration horizon direction. The above boundary condition thus physically means that we are imposing an ingoing boundary condition at the outer horizon $r=r_+$ (i.e. wave can only fall into the black hole) and an outgoing boundary condition at the acceleration horizon $r=\frac{1}{\alpha}$ (i.e. the wave can only fade out to the causally separated region beyond the acceleration horizon), see Figure \ref{fig:boundarycondition}. This is also the most natural boundary condition to be considered in an isolated black hole system \footnote{If $\alpha \to 0$ which is the black hole without acceleration, the corresponding boundary condition is the ingoing boundary condition at the horizon and the outgoing to the infinity.}.  

The QNM $\omega$ of the charged C-metric has been numerically studied in \cite{Destounis:2020pjk}, and they are classified into three families: photon surface (PS) modes $\omega_{\rm PS}$, acceleration modes $\omega_\alpha$, and near extremal modes $\omega_{\rm NS}$. The PS modes, related to the light rays on the photo surface around the black hole, are complex in general, while the latter two families of modes are observed to be purely imaginary. It is numerically found \cite{Destounis:2020pjk} that when $\alpha$ is small, the accelerating modes can be approximated by 
 \begin{equation}
 \omega_{\alpha}\sim -\alpha\lt(m-m_0+\ell+n+1\rt),\quad n=0,1,2,\dots\label{omega-alpha}
 \end{equation}
The near extremal modes only become significant when $Q$ approaches $M$, or equivalently $r_-\sim r_+$, and in this region, it is approximately given by 
  \begin{equation}
 \omega_{NE}\sim -\frac{|f'(r_+)|}{2}\lt(m-m_0+\ell+n+1\rt),\quad n=0,1,2,\dots,\label{omega-NE}
 \end{equation}
 where $\kappa_\pm=\frac{|f'(r_\pm)|}{2}$ is the surface gravity at the outer/inner horizon $r_\pm$. 

Some concrete numerical results obtained from the {\it Mathematica} package {\it QNMspectral} developed in \cite{Jansen:2017oag} are shown in Table \ref{t:QNM1} and \ref{t:QNM2} (more numerical results can be found in \cite{Destounis:2020pjk}) . In this work, we solve for these QNM by mapping both the radial and the angular equations \eqref{eq:BHODE-r} and \eqref{eq:BHODE-theta} to Heun-type equations with the help of the instanton partition functions. 

\begin{table}
    \centering
    \begin{tabular}{|c|c|c|c|}
    \hline
       & $\alpha M=0.05$ & $\alpha M=0.1$ & $\alpha M=0.1$ \\
       & $Q/M=0.3$ & $Q/M=0.3$ & $Q/M=0.8$ \\
    \hline
      $\ell=0$ & $\lambda=0.331738$ & $\lambda=0.326900$ & $\lambda=0.328775$  \\
      $m_0=0$ & $\omega_{{\rm PS}_1}=0.111-0.104i$ & $\omega_{{\rm PS}_1}=0.1079-0.1012i$ &  $\omega_{{\rm PS}_1}=0.122970-0.102222i$ \\
    & $\omega_{\alpha_1}=-0.0506i$ & $\omega_{\alpha_1}=-0.1055i$ & $\omega_{\alpha_1}=-0.104288i$ \\
    & $\omega_{\alpha_2}=-0.103i$ & $\omega_{\alpha_2}=-0.215i$ & $\omega_{\alpha_2}=-0.21i$ \\
    \hline
    $\ell=2$ & $\lambda=7.375460$ & $\lambda=8.498292$ & $\lambda=8.557553$\\
    $m_0=2$ & $\omega_{{\rm PS}_1}=0.5246-0.0967i$ & $\omega_{{\rm PS}_1}=0.5454-0.09535i$ & $\omega_{{\rm PS}_1}=0.623833- 0.0970593i$  \\
    & $\omega_{\alpha_1}=-0.16i$ & $\omega_{{\rm PS}_2}=0.5316-0.2886i$ & $\omega_{{\rm PS}_2}=0.6121-0.2933i$\\
    & & $\omega_{\alpha_1}=-0.344i$ & $\omega_{\alpha_1}=-0.34i$ \\
    \hline
    \end{tabular}
    \caption{Some example numerical data obtained from the {\it QNMspectral} package, and the computation of $\lambda$ and the explanation of $\ell$ can be found in Appendix \ref{a:angular}.}
    \label{t:QNM1}
\end{table}

\begin{table}
    \centering
    \begin{tabular}{|c|c|c|c|}
    \hline
       & $\alpha M=0.3$ & $\alpha M=0.5$ & $\alpha M=0.3$ \\
       & $Q/M=0.8$ & $Q/M=0.8$ & $Q/M=0.999$\\
       \hline
       $\ell=0$ & $\lambda=0.290224$ & $\lambda=0.194511$ & $\lambda=0.303261$\\
       $m_0=0$ & $\omega_{{\rm PS}_1}=0.09664-0.08271i$ & $\omega_{{\rm PS}_1}=0.03945-0.04122i$ & $\omega_{{\rm PS}_1}=0.11168-0.08142i$\\
       & $\omega_{{\rm PS}_2}=0.0781-0.2519i$ & $\omega_{{\rm PS}_2}=0.03876-0.1237i$ & $\omega_{{\rm NE}_1}=-0.0412i$ \\
       & $\omega_{\alpha_1}=-0.35i$ & & $\omega_{{\rm NE}_2}=-0.084i$ \\
    \hline
    $\ell=1$ & $\lambda=4.773332$  & $-$ & $\lambda=4.897025$ \\
    $m_0=1$ & $\omega_{\rm PS_1}=0.3571-0.08265i$ & & $\omega_{\rm PS_1}=0.460-0.078i$ \\
     & $\omega_{\rm PS_2}=0.3532-0.2482i$ & & $\omega_{\rm PS_2}=0.45-0.24i$\\
      & & & $\omega_{\rm NE_1}=-0.115i$\\
      \hline
    \end{tabular}
    \caption{Some example numerical data obtained from the {\it QNMspectral} package. As the value of $\lambda$ could not be obtained from the package (see Appendix \ref{a:angular}), its modes are not computed here. }
    \label{t:QNM2}
\end{table}

\subsection{Dictionary between BH data and SW data}\label{s:dic}

To solve the radial part of the QNM equation \eqref{eq:BHODE-r}, 
\begin{equation}
    \frac{{\rm d}^2\phi(r)}{{\rm d}r^2_\ast}+(\omega^2-V_r)\phi(r)=0,
\end{equation}
whose singularities are all regular and locate at $\pm\frac{1}{\alpha}, r_\pm=M\pm \sqrt{M^2-Q^2}$ when brought to the Schr\"odinger equation, from the gauge theory approach, we map its four singularities into $0$, $1$, $t$ and $\infty$ to match with the BPZ equation in the semiclassical limit \eqref{eq:SW-eq}. The latter equation will be called the SW ODE later on in this article. There are different ways to construct such maps, but in this article, we focus on two particular dictionaries\footnote{As will be explained in section \ref{s:connf}, the connection formula we use to solve the boundary conditions is based on the instanton expansion in terms of small parameter $t$. 
The parameter $t$ in these two dictionaries remains perturbative to code different parameter domains of the black hole, which enables us to apply two simple versions of the connection formulae. } which respectively map $r=r_+$ and $r=\frac{1}{\alpha}$ into $z=t,1$ and $z=0,t$. 

\subsubsection*{Dictionary 1}

The first dictionary that maps \eqref{eq:BHODE-r} to the SW ODE \eqref{eq:SW-eq} is realized by
\begin{equation}
    \psi(z)=\lt(\lt(r_--\frac{1}{\alpha}\rt)z+\frac{2}{\alpha}\rt)\sqrt{f(r)}\phi(r),\quad z=-\frac{2\alpha(r-r_-)}{(\alpha r_--1)(\alpha r+1)},
\end{equation}
under the dictionary, 
\begin{equation}\label{eq:dic-1}
    \begin{aligned}
     &t=\frac{2\alpha\left(r_{-}-r_{+}\right)}{\left(\alpha r_{-}-1\right)\left(\alpha r_{+}+1\right)},\\
    &u=-\frac{\left(\alpha r_{+}-1\right){}^{2}\left(-6\lambda+3\alpha r_{+}+\alpha r_{-}\left(\alpha r_{+}-3\right)-1\right)+\frac{12r_{+}^{3}\omega^{2}\left(r_{-}\left(\alpha r_{+}-2\right)+r_{+}\right)}{\left(r_{-}-r_{+}\right){}^{2}\left(\alpha r_{+}+1\right)}}{6\left(\alpha r_{-}+1\right)\left(\alpha r_{+}-1\right){}^{3}},\\
    &a_{0}=\frac{ir_{-}^{2}\omega}{\left(r_{+}-r_{-}\right)\left(1-\alpha^{2}r_{-}^{2}\right)},\quad a_{1}=\frac{i\omega}{2\alpha\left(1-\alpha r_{-}\right)\left(1-\alpha r_{+}\right)},\\
    &a_{t}=\frac{ir_{+}^{2}\omega}{\left(r_{+}-r_{-}\right)\left(1-\alpha^{2}r_{+}^{2}\right)},\quad a_{\infty}=\frac{i\omega}{2\alpha\left(\alpha r_{-}+1\right)\left(\alpha r_{+}+1\right)},
    \end{aligned}
\end{equation}
where we have chosen the signs of $a_i$ that do not affect the final results.
Note that the outer horizon and the boundary $r=\alpha^{-1}$ are respectively mapped to 
\begin{equation}
    r=\frac{1}{\alpha}\,\leftrightarrow\, z=1,\qquad   r=r_+\,\leftrightarrow\, z=t,
\end{equation}
so the boundary conditions \eqref{eq:bdy-BH} become
\begin{equation}\label{eq:bdy-SW}
    \psi(z)\sim\begin{cases}
(t-z)^{\frac{1}{2}-a_{t}} & z\to t\\
(1-z)^{\frac{1}{2}-a_{1}} & z\to1
\end{cases}\,,
\end{equation}
in the $z$-coordinate.

\begin{table}
    \centering
    \begin{tabular}{|c|c|c|c|c|c|c|}
        \hline
        & $\alpha M=0.05$ & $\alpha M=0.1$ & $\alpha M=0.1$ & $\alpha M=0.3$ & $\alpha M=0.5$ & $\alpha M=0.3$ \\
       & $Q/M=0.3$ & $Q/M=0.3$ & $Q/M=0.8$ & $Q/M=0.8$ & $Q/M=0.8$ & $Q/M=0.999$ \\
       \hline
       $t$ & $0.174209$ & $0.320682$ & $0.215517$ & $0.552826$ & $0.833333$ & $0.0572592$  \\
       \hline
    \end{tabular}
    \caption{The values of $t$ evaluated in dictionary 1 for different parameters.}
    \label{tab:t-value1}
\end{table}

\subsubsection*{Dictionary 2}

The second map we consider is realized by the coordinate transformation, 
\begin{equation}
    z=\frac{1+\alpha r_-}{1+\alpha r_+}\frac{r-r_+}{r-r_-},\label{map-2}
\end{equation}
with the identifications, 
\begin{align}\label{eq:map-ai-BHrpm}
    &t=\frac{(1-\alpha r_+)(1+\alpha r_-)}{(1+\alpha r_+)(1-\alpha r_-)},\\
    &u=\frac{\lambda}{2(r_+-r_-)\alpha}+\frac{(1-3r_+\alpha)\omega^2}{4(r_+-r_-)\alpha^3(1-r_-\alpha)(1-r_+\alpha)^2}+\frac{1-3r_+\alpha+r_-\alpha(3-r_+\alpha)}{12\alpha(r_+-r_-)},\cr
    &a_0=\frac{ir_+^2\omega}{(r_+-r_-)(1-r_+^2\alpha^2)},\quad a_1=\frac{i\omega}{2\alpha(1+r_-\alpha)(1+r_+\alpha)},\cr
    &a_t=\frac{i\omega}{2\alpha(1-r_-\alpha)(1-r_+\alpha)},\quad a_\infty=\frac{ir_-^2\omega}{(r_+-r_-)(1-r_-^2\alpha^2)}.
\end{align}
In this dictionary, the boundaries are mapped to 
\begin{equation}
    r=r_+\leftrightarrow z=0,\quad r=\frac{1}{\alpha}\leftrightarrow z=t.
\end{equation}

\begin{table}
    \centering
    \begin{tabular}{|c|c|c|c|c|c|c|}
    \hline
        & $\alpha M=0.05$ & $\alpha M=0.1$ & $\alpha M=0.1$ & $\alpha M=0.3$ & $\alpha M=0.5$ & $\alpha M=0.3$ \\
       & $Q/M=0.3$ & $Q/M=0.3$ & $Q/M=0.8$ & $Q/M=0.8$ & $Q/M=0.8$ & $Q/M=0.999$ \\
       \hline
       $t$ & $0.825791$ & $0.679318$ & $0.784483$ & $0.447174$ & $0.166667$ & $0.942741$  \\
       \hline
    \end{tabular}
    \caption{The values of $t$ evaluated in dictionary 2 for different parameters.}
    \label{tab:t-value2}
\end{table}

\subsection{An apparent puzzle in the B-cycle quantization}

As mentioned in section \ref{s:WKB}, one can perform an almost parallel calculation to that presented in \cite{Aminov:2020yma} to use the quantization condition of the B-cycle to compute the QNM of the charged C-metric from the instanton partition function of 4d SU(2) gauge theory with $N_f=4$ flavors. 

The B-cycle quantization condition that is expected to realize the boundary condition \eqref{eq:bdy-SW} in our convention is given by, 
\begin{equation}
    \Pi_B^{N_f=4}\lt(u(\omega),\boldsymbol{m}(\omega),\mathfrak{q}(\omega),1\rt)=\pi i\lt(2n+1\rt),\label{B-quant}
\end{equation}
for non-negative integers $n$, and the explicit form of the quantum B-period $\Pi_B$ is given by the instanton partition function as (see Appendix \ref{a:Nekra} for a brief derivation), 
\begin{align}
    \Pi^{N_f=4}_B(u,\boldsymbol{m},\mathfrak{q},\hbar)=\pi i\hbar -2\hbar \log\frac{\Gamma\lt(1+\frac{2a}{\hbar}\rt)}{\Gamma\lt(1-\frac{2a}{\hbar}\rt)}-i\hbar \sum_{j=1}^4\log\frac{\Gamma\lt(1-\frac{m_j+a}{\hbar}\rt)}{\Gamma\lt(1-\frac{m_j-a}{\hbar}\rt)}\cr
    +\partial_aF^{N_f=4}_{\rm inst}(a,\boldsymbol{m},\mathfrak{q},\hbar),
\end{align}
where the NS free energy is related to the instanton partition function $Z^{(N_f=4)}_{inst}$ via 
\begin{equation}
    F^{N_f=4}_{\rm inst}(a,\boldsymbol{m},\mathfrak{q},\hbar)=-\hbar\lim_{\epsilon_2\rightarrow 0}\epsilon_2\log Z^{(N_f=4)}_{\rm inst}(a,\boldsymbol{m},\hbar,\epsilon_2).\label{F-NS}
\end{equation}
An apparent puzzle here is that in contrast to the existence of three families of QNM in the charged C-metric, there is only one principle quantum number $n$ in the quantization condition. Then it is very curious how to reproduce different families of QNM from only one quantum number. Indeed, after computing the B-cycle quantization condition \eqref{B-quant} for the charged C-metric, only complex QNM identified as the PS modes are obtained, which is consistent with the picture of the QNM as resonant states and that the PS modes are usually the family obtained in the WKB approximation\footnote{One can similarly check that only PS modes can be obtained from the Bender-Wu approach proposed in \cite{Hatsuda:2019eoj}.}. For example at 3-instanton level, we obtain $\omega=0.106716-0.0986724i$ in the case of $\alpha M=0.1$, $Q/M=0.3$, $\lambda=0.326900$, $n=0$, compared with the numerical result of the PS mode $\omega_{{\rm PS}_1}=0.1079-0.1012i$, while it is very subtle in this approach how to get purely imaginary modes. In this article, we try to resolve this apparent puzzle by showing that implementing the connection-formula approach described in the next section instead of the B-cycle quantization condition improves the algorithm in finding the QNM and reproduces all the families of the modes in a clear manner.

\section{Connection formula approach to QNM}\label{s:connf}

Even though we mentioned that the Heun equation can be solved in terms of the instanton partition function in the Nekrasov-Shatashvili limit, it is still unfortunately given as a power series expansion, and therefore is only valid inside its convergence radius. In the case of solving QNM of the charged C-metric, we need to impose the boundary condition at $r=r_+$ and $r=\frac{1}{\alpha}$, and around these two end-points, the general solution is given respectively by the linear combination of two independent solutions (in the form of series expansions). In dictionary 1, these boundaries locate at $z=t$ and $z=1$, and the general solution at $z=t$ is given by  
\begin{align}
    \psi(z)=C_1\psi^{(t)+}(z)+C_2\psi^{(t)-}(z),
\end{align}
where $\psi^{(t)\pm}$ are series expansions defined around $z\sim t$, and similarly 
\begin{align}
    \psi(z)=D_1\psi^{(1)+}(z)+D_2\psi^{(1)-}(z),
\end{align}
holds around $z\sim 1$ for some constants $C_{1,2}$ and $D_{1,2}$ chosen properly to satisfy the boundary conditions. If we know how the solutions $\psi^{(t)\pm}$ are analytically continued to the region around $z\sim 1$, 
\begin{equation}
    \psi^{(t)\pm}(z)=\sum_{\theta=\pm}A_{\pm,\theta}\psi^{(1)\theta}(z),\label{scheme-CF}
\end{equation}
we will be able to relate the constants $C_{1,2}$ and $D_{1,2}$ used on two boundaries, 
\begin{equation}
    \frac{D_1}{D_2}=\frac{C_1A_{++}+C_2A_{-+}}{C_1A_{+-}+C_2A_{--}}.
\end{equation}
As there is $\omega$-dependence in $A_{\theta,\theta'}$, the above equation allows one to solve for the QNM $\omega$ rigidly given the precise expressions of $A_{\theta,\theta'}$. 

We call the analytic continuation \eqref{scheme-CF} the connection formula\footnote{The most well-known example of the connection formula is the one in the context of the hypergeometric function, 
\begin{eqnarray}\label{CF-hyperg}
    {}_2F_1(a,b,c;y)&=&\frac{\Gamma(c)\Gamma(c-a-b)}{\Gamma(c-a)\Gamma(c-b)}{}_2F_1(a,b,1+a+b-c;1-y) \\
    &&+\frac{\Gamma(c)\Gamma(a+b-c)}{\Gamma(a)\Gamma(b)}(1-y)^{c-a-b}{}_2F_1(c-a,c-b,1+c-a-b;1-y). \nonumber
\end{eqnarray}
It connects the solution around one singular point $y=0$ and those around another one at $y=1$. }. 
In our work, we need to work out such connection formulae for the solutions to the Heun equation, i.e. Heun functions. 
We present the connection formulae for the Heun functions in this section, and the derivations of part of these formulae are based on the perspective of 2d CFT as enlightened by \cite{Bonelli:2022ten}.
More details are listed in
Appendix \ref{a:conn-form}. 
The connection formulae will be shown to reproduce all known families of QNM in the charged C-metric. 
We will conclude by briefly discussing the difference between the B-cycle quantization condition
given in \eqref{B-quant} and the connection formula used in this section. 

\subsection{Connection formula and numerical results}

In this section, we perform the analysis on the QNM with the connection formulae. In different dictionaries, the boundaries are mapped to different positions, so we need to use different connection formulae. In the following, we present the numerical analysis respectively in dictionary 1 and 2 described in section \ref{s:dic}. 

\subsubsection*{Dictionary 1}

\begin{figure}
    \centering
    \includegraphics[width=12cm]{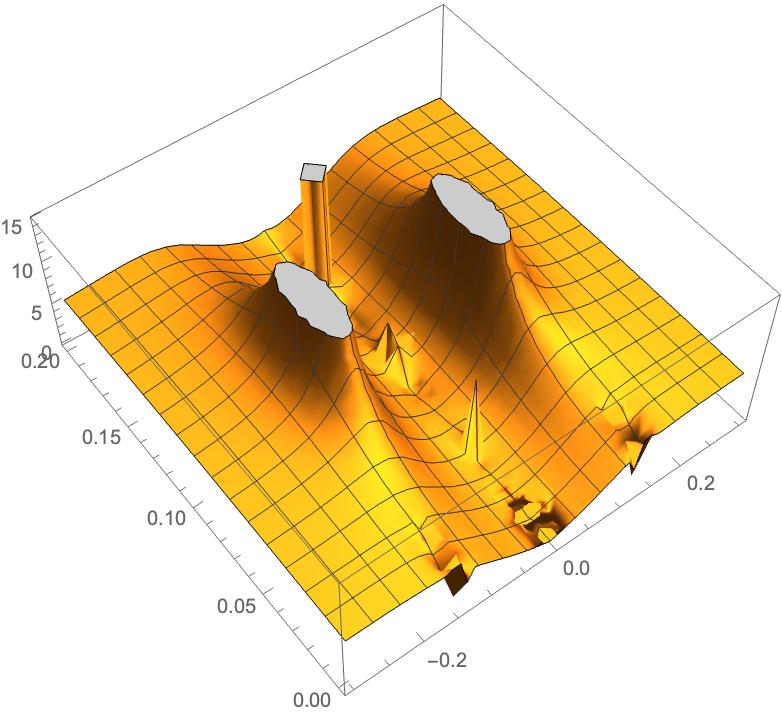}
    \caption{The plot of $1/|A_{-+}|$ on the plane of $\omega=x-iy$ including 3-instanton corrections with the spikes (poles) corresponding to QNM in the charged C-metric.}
    \label{fig:conn-plot}
\end{figure}

One can solve \eqref{eq:SW-eq} explicitly around $z\sim t$ to obtain two independent solutions $\psi^{(t)\pm}$ as series expansions (see Appendix \ref{a:Heun} for more details), and we see that they respectively behave asymptotically as, 
\begin{equation}
    \psi^{(t)\pm}(z)\sim (t-z)^{\frac{1}{2}\pm a_t},\quad z\sim t.
\end{equation}
That is to say $\psi^{(t)-}$ is exactly the ingoing solution that satisfies the boundary condition \eqref{eq:bdy-SW}, and we need to continue it to the other boundary $z\sim 1$. Similarly the independent solutions to the SW ODE \eqref{eq:SW-eq} at boundary $z\sim 1$ behave asymptotically as
\begin{equation}
    \psi^{(1)\pm}\sim (1-z)^{\frac{1}{2}\pm a_1}.
\end{equation}
The solution at the outer horizon $z\sim t$ (corresponding to $r\sim r_+$) and the ones at the boundary $z\sim 1$ (corresponding to $r\sim \frac{1}{\alpha}$) are related by
\begin{equation}
    \psi^{(t)-}(z)=\sum_{\theta^{\prime}=\pm}A_{-\theta^{\prime}}\psi^{(1)\theta^{\prime}}(z),
\end{equation}
where the coefficients $A_{-\pm}$ are given by \cite{Bonelli:2022ten,Dodelson:2022yvn}
\begin{align}
A_{-\theta^{\prime}}=\lt(\sum_{\sigma=\pm}{\cal M}_{-\sigma}(a_{t},a;a_{0}){\cal M}_{(-\sigma)\theta^{\prime}}(a,a_{1};a_{\infty})t^{\sigma a}e^{-\frac{\sigma}{2}\partial_{a}F}\rt)t^{\frac{1}{2}-a_{0}-a_{t}}(1-t)^{a_{t}-a_{1}}\cr
\times e^{-\frac{1}{2}(\partial_{t}+\theta^{\prime}\partial_{a_{1}})F},\label{eq:coeff}
\end{align}
and it can be derived by combining the connection formulae \eqref{conn-w-0t}, \eqref{conn-w-0inf} and \eqref{conn-w-1inf}. 
Here $F$ is the NS free energy of the instanton partition function \eqref{F-NS}, and ${\cal M}$ is defined by the $\Gamma$-function as \cite{Bonelli:2022ten}
\begin{equation}
    {\cal M}_{\theta\theta^{\prime}}(\alpha_{0},\alpha_{1};\alpha_{2})=\frac{\Gamma(-2\theta^{\prime}\alpha_{1})}{\Gamma(\frac{1}{2}+\theta\alpha_{0}-\theta^{\prime}\alpha_{1}+\alpha_{2})}\frac{\Gamma(1+2\theta\alpha_{0})}{\Gamma(\frac{1}{2}+\theta\alpha_{0}-\theta^{\prime}\alpha_{1}-\alpha_{2})}.\label{M-def}
\end{equation}
To satisfy the boundary condition \eqref{eq:bdy-SW}, we should impose the following condition
\begin{equation}
    A_{-+}=0.\label{conn-form1}
\end{equation}
More explicitly it says 
\begin{align}
    \lt[\cM_{-+}(a_t,a;a_0)\cM_{-+}(a,a_1;a_\infty)t^ae^{-\frac{1}{2}\partial_aF}+\cM_{--}(a_t,a;a_0)\cM_{++}(a,a_1;a_\infty)t^{-a}e^{\frac{1}{2}\partial_aF}\rt]\cr
    \times t^{\frac{1}{2}-a_{0}-a_{t}}(1-t)^{a_{t}-a_{1}}e^{-\frac{1}{2}(\partial_{t}+\partial_{a_{1}})F}=0.\label{explicit-CF}
\end{align}
We solved the above equation numerically for different sets of parameters and the results are shown in Table \ref{t:CF-1}-\ref{t:CF-4}. From the plot of $1/A_{-+}$ (Figure \ref{fig:conn-plot}), one can visually see that not only the photon surface modes are reproduced, but the purely imaginary accelerating modes are also obtained from the connection formula analysis. We observe from the numerical data that when $t$ is small, the numerical results computed from the connection formula including even a few instanton corrections reproduce the data generated from the {\it QNMspectral} package at very good accuracy (e.g. for $\alpha M=0.05, Q/M=0.3\Rightarrow t=0.174209$ and $\alpha M=0.3, Q/M=0.999\Rightarrow t=0.0572592$). This is a very natural result, as both the wavefunction and the connection formula between solutions in different regions are given in terms of the instanton partition function expanded in terms of the instanton counting parameter $\mathfrak{q}=t$ in the current approach. 

\paragraph{Remark:} We note that only a few number of instanton corrections have been computed in this article and mathematically we cannot conclude that the precision of the five-instanton result is better than the three-instanton one. The precision of the numerical data, for example in Table \ref{t:CF-1}, is also not good enough to make such a comparison. One should also be careful that for example in {\bf Dictionary 1} \eqref{eq:dic-1}, the instanton counting parameter $t\propto \alpha$, and one would naturally expect the instanton expansion converges better in the small-$\alpha$ region. The parameters $a_1$ and $a_\infty$ (respectively related to the surface gravity at $r=\pm \frac{1}{\alpha}$), however, are proportional to $\frac{1}{\alpha}$, and the expansion coefficients may grow quickly when $\alpha$ is approaching zero\footnote{We would like to thank one of the referees for pointing out this.}, which compete with the expansion parameter $t$ to determine the convergence of the full instanton series. We plan to discuss about the convergence and the precision of the instanton series with a more efficient program to be presented in our future work.

\begin{table}
    \centering
    \begin{tabular}{|c|c|c|c|}\hline
      &  3-instanton  &  5-instanton & Numerical data   \\\hline
      $\omega_{\alpha_1}$&   $-0.0505002 i$&$-0.0505526i$ & $- 0.0506 i$  \\
      $\omega_{\alpha_2}$&   $-0.10341 i$& $-0.103672i$ & $-0.103i$  \\
      \hline
      $\omega_{\rm PS_1}$&$0.111416 - 0.10266i$& $0.112086-0.10318i$&$0.111-0.104i$\\
      $\omega_{\rm PS_2}$&$0.097464 - 0.33332 i$& &\\
      \hline
    \end{tabular}
    \caption{Results obtained from the connection formula at $\alpha M=0.05, Q=0.3 M, \lambda=0.3317$, $t=0.174209$, compared with numerical data obtained from the {\it QNMspectral} package. We also included the result predicted for $\omega_{\rm PS_2}$ at 3-instanton in the above table. In the {\it QNMspectral} package, we could not extract out the numerical information for this mode due to the limitation of the numerical approach. We also remark that it seems to be in general easier in the connection formula approach to analyze the behavior of higher modes. }
    \label{t:CF-1}
\end{table}


\begin{table}
    \centering
    \begin{tabular}{|c|c|c|c|}\hline
      & 3-instanton & 5-instanton  & Numerical data \\\hline
      $\omega_{\rm NE_1}$&   $-0.0412224i$& $-0.0412147i$ & $- 0.0412 i$  \\
      $\omega_{\rm NE_2}$& $-0.0837135i$ & $-0.0836822i$ & $-0.084i$ \\
      $\omega_{\rm PS}$&$0.111516-0.0839528i$ & $0.112347-0.0844511 i$
      & $0.11168-0.08142i$\\\hline
    \end{tabular}
    \caption{Results obtained from the connection formula at $\alpha M=0.3, Q=0.999 M,\lambda=0.3033$, $t=0.0572592$, compared with numerical data obtained from the {\it QNMspectral} package.}
    \label{t:CF-2}
\end{table}

\begin{table}
    \centering
    \begin{tabular}{|c|c|c|c|}\hline
      & 3-instanton & 5-instanton  & Numerical data  \\\hline
      $\alpha=0.1,Q=0.3$ & $\omega_{\alpha_1}=-0.107039i$ & $-0.106378i$ & $-0.1055i$ \\
      $\lambda=0.326900$ & $\omega_{\rm PS_1}=0.0991-0.0950i$ & $0.1003-0.0965i$ & $0.1079-0.1012i$  \\
      $t=0.320682$ & & & \\
      \hline
      $\alpha=0.5,Q=0.8$ & $\omega_{\alpha_1}=0.0338-0.0335i$ & $0.0348-0.0364i$ & $0.03945-0.04122i$ \\
      $\lambda=0.194511$ & $\omega_{\rm PS_1}=0.0573-0.0986i$ & $0.0515-0.1078i$ & $0.03876-0.1237i$  \\
      $t=0.833333$ & & & \\
      \hline
    \end{tabular}
    \caption{Comparison of results obtained at 3-instanton and 5-instanton level with numerical data obtained from the {\it QNMspectral} package for different parameters with various $t$. $M$ is set to 1 here without losing any generality. }
    \label{t:CF-3}
\end{table}

\begin{table}
    \centering
    \begin{tabular}{|c|c|c|}\hline
      & 3-instanton & Numerical data \\\hline
      $\alpha=0.1,Q=0.8$ & $\omega_{\alpha_1}=-0.101239i$ & $-0.104288i$ \\
      $\lambda=0.328775$ & $\omega_{\alpha_2}=-0.209259i$ & $-0.21i$  \\
      $t= 0.215517$ & $\omega_{\rm PS_1}=0.121139-0.101221i$ & $0.122970-0.102222i$  \\
      \hline
      $\alpha=0.3,Q=0.8$ & $\omega_{\alpha_1}=-0.414573i$ & $-0.35i$ \\
      $\lambda=0.290224$ & $\omega_{\rm PS_1}=0.0899294-0.0781695i$ & $0.09664-0.08271i$ \\
      $t=0.552826$ & $\omega_{\rm PS_2}=0.089161-0.236303i$ & $0.0781-0.2519i$ \\
      \hline
      $\alpha=0.05,Q=0.3$ & $\omega_{\alpha_1}=-0.160143i$ & $-0.16i$ \\
      $\lambda=7.375460 $ & $\omega_{\rm PS_1}=0.526102-0.0955259i$ & $0.5246-0.0967i$  \\
      $0.174209$ & & \\
      \hline
      $\alpha=0.3,Q=0.999$ & $\omega_{\rm NE_1}=-0.114686i$ & $-0.115i$ \\
      $\lambda=4.897025 $ & $\omega_{\rm PS_1}=0.466471-0.0760236i$ & $0.460-0.078i$ \\
      $t=0.0572592$ & $\omega_{\rm PS_2}=0.45288-0.244597i$ & $0.45-0.24i$ \\
      \hline
    \end{tabular}
    \caption{Comparison of QNM obtained at 3-instanton with numerical data obtained from the {\it QNMspectral} package for different parameters with various $t$. $M$ is set to 1 here without losing any generality.}
    \label{t:CF-4}
\end{table}

\subsubsection*{Dictionary 2}

Dictionary 2 given by \eqref{map-2} can partially compensate the numerical lists obtained from dictionary 1. As one can see from Table \ref{tab:t-value2}, for example for $\alpha M=0.5$, $Q/M=0.8$, dictionary 2 gives a relatively small $t=0.166667$ in contrast to its value $t=0.833333$ in dictionary 1. In dictionary 2, one boundary is mapped to $z=0$ and the other is mapped to $z=t$, so we need to use the connection formula shown in \eqref{conn-w-0t}, and one can require the $\Gamma$-function part to be zero with instanton corrections included through the Matone relation \cite{Matone:1995rx} discussed in Appendix \ref{a:Matone}. 
In Table \ref{t:CF-5}, we show that indeed for small $t$, the numerical results obtained from the connection formula in dictionary 2 with only 3-instanton corrections reproduce the results calculated from the {\it QNMspectral} package in very good accuracy. 


\begin{table}
    \centering
    \begin{tabular}{|c|c|c|}\hline
      & 3-instanton & Numerical data \\\hline
      $\alpha M=0.5,Q/M=0.8$ & $\omega_{\rm PS_1}=0.0394654-0.0412316i$ & $0.03945-0.04122i$ \\
      $\lambda=0.194511$ & $\omega_{\rm PS_2}=0.0387606-0.123697i$ & $0.03876-0.1237i$ \\
      $t=0.166667$ & &\\
      \hline
      $\alpha M=0.3,Q/M=0.8$ & $\omega_{\rm PS_1}=0.0979856-0.083472i$ & $0.09664-0.08271i$ \\
      $\lambda=0.290224$ & $\omega_{\rm PS_2}=0.0766748-0.252914i$ & $0.0781-0.2519i$  \\
      $t=0.447174$ & $\omega_{\alpha_1}=-0.343135i$ & $-0.35i$ \\
      \hline
      \end{tabular}
    \caption{Comparison of QNM obtained at 3-instanton with numerical data obtained from the {\it QNMspectral} package for different parameters with various $t$. $M$ is set to 1 here without losing any generality.}
    \label{t:CF-5}
\end{table}

\subsection{Analytical results on QNM}

One main advantage of the quantized Seiberg-Witten approach to the QNM is that one can obtain analytical results of the QNM based on the expansion over some small parameters. It is useful to rewrite the connection formula in dictionary 1 as
\begin{equation}\label{conn-form1-re}
    \begin{aligned}
    &t^{\frac{1}{2}-a_{0}-a_{t}}(1-t)^{a_{t}-a_{1}}e^{-\frac{1}{2}(\partial_{t}+\partial_{a_{1}})F}t^{a}e^{-\frac{1}{2}\partial_{a}F}{\cal M}_{-+}(a_{t},a;a_{0}){\cal M}_{-+}(a,a_{1};a_{\infty})\\&\qquad\qquad\qquad\qquad \times\Big(1+\frac{{\cal M}_{--}(a_{t},a;a_{0}){\cal M}_{++}(a,a_{1};a_{\infty})}{{\cal M}_{-+}(a_{t},a;a_{0}){\cal M}_{-+}(a,a_{1};a_{\infty})}t^{-2a}e^{\partial_{a}F}\Big)=0.
    \end{aligned}
\end{equation}

\subsubsection{QNM in small $\alpha M$ region}\label{s:ana-QNM1}
Let us consider the case with $\alpha M\ll 1$ very small, where the instanton counting parameter in dictionary 1 can be expanded with the leading-order behavior, 
\begin{equation}
    t= 4\sqrt{M^2-Q^2}\alpha+\cdots,
\end{equation}
such that the instanton corrections can be ignored at the leading order. Taking the leading order of the Matone relation \eqref{Matone-u-a}, we obtain 
\begin{equation}
    a=\sqrt{a_0^2+a_t^2-u-1/4}+\cdots=\sqrt{\lambda-\frac{1}{12}+2\omega^{2}\big(Q^{2}-2M(\sqrt{M^{2}-Q^{2}}+2M)\big)}+\cdots
\end{equation}
 The accelerating modes $\omega_\alpha$ turn out to come from the overall factor $\frac{1}{\Gamma(\frac{1}{2}-a-a_1-a_\infty)}$ in the connection formula \eqref{conn-form1-re}, whose zeroes are nothing but the poles of the Gamma function.
One thus finds the level-$n$ accelerating mode as an expansion with respect to $\alpha$ and the leading-order behavior is given by 
\begin{equation}
    \omega_{\alpha_n}= -i \alpha\lt(\frac{1}{2}+n+\sqrt{\lambda-\frac{1}{12}}\rt)+\cdots
\end{equation}
By further using the fact that at the leading order, $\lambda= \ell(\ell+1)+1/3+{\cal O}(\alpha)$, one obtains $\omega =-i\alpha(\ell+n+1)+\cdots$ and it reproduces the leading-order behavior of \eqref{omega-alpha} as $m\rightarrow m_0$ ($P(\pi)\rightarrow 1$) when $\alpha\rightarrow 0$. We note that one can work harder to obtain the analytic expressions for $\omega_{\alpha_n}$ as series expansions in $\alpha$.

\subsubsection{QNM in the near-extremal region} \label{s:ana-QNM2}
In the near-extremal region $r_+-r_-=:\delta r\approx 0$ of dictionary 1, which can be realized by taking $Q$ very closed to $M$, we also have 
\begin{equation}
    t= \frac{4  \alpha  \sqrt{2Q (M-Q)}}{1-\alpha
   ^2 Q^2}+\cdots,
\end{equation}
as a small expansion parameter and thus we can ignore the instanton corrections at the leading order. The Matone relation \eqref{Matone-u-a} at the leading order gives 
\begin{equation}
    a=\sqrt{\frac{\lambda}{1-\alpha^{2}Q^{2}}-\frac{1}{12}-\frac{2Q^{2}\omega^{2}\big(3+\alpha Q(\alpha Q+2)\big)}{\left(\alpha^{2}Q^{2}-1\right)^{4}}}+\cdots.
\end{equation}
The near extremal mode is due to the factor $\frac{1}{\Gamma(-a_t-a-a_0+\frac{1}{2})}$ in the connection formula \eqref{conn-form1}, whose zeroes lead to the level-$n$ near extremal mode,
\begin{align}
    \omega_{{\rm NE}_n}=& -\frac{i\sqrt{M-Q}\left(1-\alpha^{2}Q^{2}\right)\left(1+2n+\sqrt{\frac{4\lambda}{1-\alpha^{2}Q^{2}}-\frac{1}{3}}\right)}{\sqrt{2}Q^{3/2}}+\cdots\\
    =& -\frac{i\sqrt{1-Q/M}\left(1-\alpha^{2}M^2\right)\left(1+2n+\sqrt{\frac{4\lambda}{1-\alpha^{2}M^2}-\frac{1}{3}}\right)}{\sqrt{2}M}+\cdots
\end{align}
Comparing with the asymptotic behavior, 
\begin{equation}
    f'(r_+)\sim -f'(r_-)\sim \frac{2\sqrt{2}}{M}(1-\alpha^2M^2)\sqrt{1-Q/M}, 
\end{equation}
we reproduce the numerical observation \eqref{omega-NE} for $\alpha M\ll 1$. Our result further predicts finite $\alpha$-corrections for the near extremal modes.

\subsection{Comparison with the B-cycle quantization condition}

We have shown strong supportive evidence that the connection formula can reproduce all families of the QNM in the charged C-metric, while the B-cycle quantization condition clearly does not. Let us rewrite the connection formula \eqref{conn-form1} into a different form to see its relation with the the B-cycle quantization condition. 

Assuming that the overall factor in \eqref{conn-form1-re} is non-zero, then the connection formula \eqref{conn-form1-re} can be re-organized into 
\begin{align}
    \frac{\cM_{--}(a_t,a;a_0)\cM_{++}(a,a_1;a_\infty)}{\cM_{-+}(a_t,a;a_0)\cM_{-+}(a,a_1;a_\infty)}t^{-2a}e^{\partial_aF}=-1.
    \label{rewrite-conn}
\end{align}
To compare this equation with the B-cycle quantization condition \eqref{B-quant} in the case $\hbar=1$, we need to first clarify the relation between the semiclassical limit taken in the conformal block and the Nekrasov-Shatashvili limit taken in the gauge theory. As one can check from the explicit expressions presented in Appendix \ref{a:Nekra}, the instanton partition function is given in terms of the combination, 
\begin{equation}
    \sigma a+\sum_{f=1}^4\sigma_fm_f+(i-1)\epsilon_1+(j-1)\epsilon_2,\quad \sigma,\sigma_f=\pm 1,\ i,j\in\mathbb{Z}_+.
\end{equation}
Under the choice $\epsilon_1=b^{-1}$ and $\epsilon_2=b$, one may factor out $b^{-1}$ from the above combination to obtain
\begin{equation}
    b^{-1}\lt(\sigma (ba)+\sum_{f=1}^4\sigma_f(bm_f)+(i-1)+(j-1)b^2\rt).
\end{equation}
In superconformal gauge theories such as the SU(2) $N_f=4$ theory considered in this article, all the prefactors $b^{-1}$ cancel out in the partition function, and the remaining factors in the limit $b\rightarrow 0$ are equivalent to those in the Nekrasov-Shatashvili limit $\epsilon_1=\hbar=1$ and $\epsilon_2=0$ with rescaled parameters, 
\begin{equation}
    \bar{a}=\lim_{b\rightarrow 0}ba,\quad \bar{m}_f=\lim_{b\rightarrow 0}m_fb.
\end{equation}

Now we recall the AGT dictionary in the semiclassical limit, 
\begin{align}
 \frac{\bar{m}_1-\bar{m}_2}{2}=a_0,\quad \frac{\bar{m}_1+\bar{m}_2}{2}-\frac{1}{2}=a_t,\\
 \frac{\bar{m}_3+\bar{m}_4}{2}-\frac{1}{2}=a_1,\quad \frac{\bar{m}_3-\bar{m}_4}{2}=a_\infty,
\end{align}
then \eqref{rewrite-conn} (with parameter $a$ in the NS limit replaced by $\bar{a}$ in the semiclassical limit) can be rewritten into 
\begin{align}
    -\frac{\Gamma(1+2\bar{a})^2}{\Gamma(1-2\bar{a})^2}\prod_{f=1}^4\frac{\Gamma(1-\bar{a}-\bar{m}_f)}{\Gamma(1+\bar{a}-\bar{m}_f)}t^{-2\bar{a}}e^{\partial_{\bar{a}}F}=-1.
\end{align}
Taking {\it logarithm} of the both sides, interestingly we recover the B-cycle quantization condition \eqref{B-quant}\footnote{During the preparation of this paper, we realized that this point is also emphasized in a recent paper \cite{Aminov:2023jve}. We also notice that $t^{-2\bar{a}}$ is related to the prefactor $\mathfrak{q}^{\Delta-\Delta_2-\Delta_1}$ appeared when the conformal bloack is expressed in terms of the instanton partition function in \eqref{eq:AGT}. One needs to be careful that there is a relative minus sign in the conventional definition \eqref{F-NS} of $F^{N_f=4}_{inst}$ to $F(t)$ in the semiclassical conformal block.}. From this, we suspect that the missing modes come from the zeroes of the overall factor in \eqref{conn-form1-re}, i.e. the assumption made to obtain \eqref{rewrite-conn} was not correct in the calculation of the QNM. This is also supported from the analytical results obtained in \ref{s:ana-QNM1} and \ref{s:ana-QNM2} based on some small-parameter expansion. The physical picture of this discrepancy in the WKB approach is yet to be explore as a future work. Non-perturbative phenomena might be responsible as the $\Gamma$-function part has a clear $\frac{1}{\hbar}$-dependence\footnote{The non-square-integrability nature of the wavefunction in this problem can also be the origin of the failure of the quantization condition, as this is an important assumption in the topological string approach to the quantization condition in 5d gauge theories \cite{Grassi:2014zfa,Wang:2015wdy}. We would like to thank Xin Wang for pointing out this possibility.}. Because log-functions are involved in the quantization condition \eqref{B-quant}, how to choose the branch cuts may also affect the final results of the computation. 

We note that one of the major advantage of the connection formula over the B-cycle quantization condition is that it avoids the ambiguity merges during the process of taking the logarithm to obtain the quantization condition. In addition, in dictionary 2, the connection formula connecting $z\sim 0$ and $z\sim t$ looks completely different from a B-cycle quantization condition, and it might not be very clear what kind of quantization condition needs to be imposed in this dictionary.

\paragraph{Remark:} One can also realize other boundary conditions with the connection formula, e.g. requiring the l.h.s. of \eqref{rewrite-conn} to blow up modifies the boundary condition at the acceleration horizon $r=\frac{1}{\alpha}$ from outgoing to ingoing. 
Moreover, the ratio of the coefficients $A_{-\theta}$ presented in \eqref{eq:coeff} $\frac{A_{--}}{A_{-+}}$ corresponds to the Green function of a scalar field defined on the charged C-metric background, and can be used to analyze for example the absorption cross section in the scattering problem. We note, however, the A-cycle in the current context does not seem to play any role, as one may naively expect the A-cycle to be related to boundary-condition problem between the outer horizon and the inner horizon.

\subsection{A similar structure in the dS black hole}

In fact, this kind of apparent puzzle is not a unique feature of the charged C-metric at all. As noted in \cite{Aminov:2020yma}, the vector and the tensor perturbation (respectively with $s=1$ and $s=2$) around the dS black hole, 
\begin{equation}
    {\rm d}s^2=-f(r){\rm d}t^2+\frac{{\rm d}r^2}{f(r)}+r^2({\rm d}\theta^2+\sin^2\theta{\rm d}\varphi^2),
\end{equation}
where 
\begin{equation}
    f(r)=1-\frac{2M}{r}-\frac{r^2}{L^2},\quad \Lambda=\frac{3}{L^2},
\end{equation}
are also dual to the SU(2) gauge theory with $N_f=4$ flavors. The e.o.m. is given by 
\begin{equation}
    \frac{{\rm d}^2\Phi}{{\rm d}r^2_\ast}+\lt(\omega^2-V_{dS}(r)\rt)\Phi=0,
\end{equation}
where ${\rm d}r_\ast=\frac{{\rm d}r}{f(r)}$ and 
\begin{equation}
    V_{\rm dS}(r)=f(r)\lt(\frac{l(l+1)}{r^2}+(1-s^2)\lt(\frac{f'(r)}{r}+\frac{s^2\Lambda}{6}\rt)\rt).
\end{equation}
There are three obvious singularities given by $f(r)=0$, and we denote them as $r_-$, $r_+$, $r_c$ with the ordering $r_-<r_+<r_c$. These singularities respectively correspond to the Cauchy horizon, the event horizon and the cosmological horizon in the dS black hole. The QNM are frequencies satisfying the boundary condition, 
\begin{equation}\label{eq:bdy-dS}
    \phi(r)\sim\begin{cases}
e^{-i\omega r_{\ast}} & r\to r_{+}\\
e^{i\omega r_{\ast}}& r\to r_c
\end{cases}.
\end{equation}
One can construct two maps similar to dictionary 1 and dictionary 2 \eqref{map-2} in the study of the charged C-metric to respectively map $r_+\rightarrow t,\ r_c\rightarrow 1$ and $r_+\rightarrow 0,\ r_c\rightarrow t$. In {\bf dictionary 1}, we use 
\begin{equation}
    z=\frac{r_c-r_-}{r_c}\frac{r}{r-r_-},
\end{equation}
with 
\begin{equation}
    t=\frac{r_+(r_c-r_-)}{r_c(r_+-r_-)}.
\end{equation}
$t$ is typically small in the region $\Lambda M^2$ is small. The potential is again mapped to the form of \eqref{QSW} with additional terms vanishing in the case $s=1,2$, 
\begin{equation}
	Q_{\rm dS}(z)=Q_{\rm SW}(z)+\frac{\Delta_s}{(z-\tilde{t})^2}+\frac{u_s}{y(y-\tilde{t})},
\end{equation}
with $\tilde{t}=1-\frac{r_-}{r_c}$, $a_0=s$, 
\begin{align}
    a_1=\frac{3ir_c\omega}{(r_c-r_-)(r_c-r_+)\Lambda},\quad u_s=-\frac{(s^2-1)(s^2-4)(r_c+r_+-r_-)}{2r_-},\\
    a_t=\frac{3ir_+\omega}{(r_c-r_+)(r_+-r_-)\Lambda},\quad \Delta_s=-\frac{1}{2}(s^2-4)(s^2-1),\\
    a_\infty=\lt(-\frac{(s^2-1)\lt(2r_-+(s^2-2)(r_c+r_+)\rt)}{2r_-}-\frac{9r_-^2\omega^2}{(r_c-r_-)^2(r_--r_+)^2\Lambda^2}\rt)^{\frac{1}{2}},
\end{align}
and 
\begin{align}
    u=\frac{6l(l+1)}{2r_-(r_c-r_+)\Lambda}+\frac{r_c(r_--r_+)(2s^2-1)-r_+(s^2-1)(2r_-+(s^2-2)r_+)}{2r_-(r_c-r_+)}\cr
    +\frac{18r_cr_+^2\omega^2}{r_-(r_c-r_+)^3(r_--r_+)\Lambda}.
\end{align}
The {\bf dictionary 2} is given by the map, 
\begin{equation}
    z=\frac{r_-}{r_--r_+}\frac{r-r_+}{r},
\end{equation}
where 
\begin{equation}
    t=\frac{r_-(r_c-r_+)}{r_c(r_--r_+)}.
\end{equation}
$t$ in this map is typically small when $\Lambda M^2$ is relatively large approaching its critical value $\Lambda M^2=\frac{1}{9}$. The potential is again mapped to the form of \eqref{QSW} with corrections vanishing in the case $s=1,2$, 
\begin{equation}
	Q_{\rm dS}(z)=Q_{\rm SW}(z)+\frac{\Delta_s}{(z-\tilde{t})^2},
\end{equation}
with the dictionary, $\tilde{t}=\frac{r_-}{r_--r_+}$, 
\begin{align}
	a_0=\frac{3ir_+\omega}{(r_c-r_+)(r_+-r_-)\Lambda},\quad a_1=\frac{3ir_-\omega}{(r_c-r_-)(r_+-r_-)\Lambda},\\
	a_t=\frac{3ir_c\omega}{(r_c-r_-)(r_c-r_+)\Lambda},\quad \Delta_s=-\frac{1}{2}(s^2-4)(s^2-1),\\ 
	a_\infty=\lt(\frac{-2r_+(1-2s^2)+r_c(s^4-3s^2+2)+r_-(s^4-3s^2+2)}{2r_+}\rt)^{\frac{1}{2}},
\end{align}
and 
\begin{align}
	u=\frac{6l(l+1)+\lt(2r_+r_-s^2+r_c(r_++r_-)(1-2s^2)-r_c^2(s^2-2)(s^2-1)\rt)\Lambda}{2(r_c-r_-)r_+\Lambda}\cr
	-\frac{18r_c^2(r_cr_++r_cr_--2r_+r_-)}{(r_c-r_-)^3(r_c-r_+)^2r_+\Lambda^2}\omega^2.
\end{align}
The dS black hole also has dS modes and Photon sphere (PS) modes as its QNM (it can be seen from the asymptotic analysis in \cite{Cardoso:2004up} for the vector and tensor perturbations and e.g. \cite{Jansen:2017oag} for the numerical results on the scalar perturbations), and it is again puzzling that one quantum number in the B-cycle quantization condition cannot reproduce two families of modes, and we again found difficulties in solving for the purely imaginary modes from the B-cycle quantization condition. This apparent puzzle, however, is still solved by the connection formula approach. In Table \ref{t:CF-6} and \ref{t:CF-7}, we show that the connection formulae \eqref{conn-form1} and \eqref{conn-w-0t} respectively in dictionary 1 and 2 reproduce the QNM computed from the {\it QNMspectral} package very well, and we observe that the precision improves when $t$ is small.  

\begin{table}
    \centering
    \begin{tabular}{|c|c|c|}\hline
      & 3-instanton & Numerical data  \\\hline
      $\Lambda M^2=0.06$ & $\omega_{\rm PS_1}=0.168616 -0.0623145i$ & $0.170891-0.063809i$ \\
      $s=l=1$ & $\omega_{\rm PS_2}=0.159303-0.188259i$ & $0.158918-0.193747i$  \\
      $t=0.523124$ & $\omega_{\rm dS_1}=-0.296564i$ & $-0.288i$  \\
      \hline
      $\Lambda M^2=0.03$ & $\omega_{\rm PS_1}=0.212917-0.0764187i$ & $0.213721-0.0795649i$ \\
      $s=l=1$ & $\omega_{\rm PS_2}=0.190808-0.245907i$ & $0.190633-0.246947i$ \\
      $t=0.360835$ & $\omega_{\rm dS_1}=-0.202359i$ & $-0.202031i$ \\
      & $\omega_{\rm dS_2}=-0.309291i$ & $-0.31i$  \\
      \hline
      $\Lambda M^2=0.005$ & $\omega_{\rm PS_1}=0.241109-0.0872755 i$ & $0.242916-0.0904662i$  \\
      $s=l=1$ & $\omega_{\rm dS_1}=-0.0818023i$ & $-0.0818024i$   \\
      $t=0.152999$ & & \\
      \hline
      $\Lambda M^2=0.01$ & $\omega_{\rm PS_1}=0.0893386-0.101808i$ & $0.1042786-0.1045020i$  \\
      $s=2,l=1$ & $\omega_{\rm dS_1}=-0.115414i$ & $-0.119142i$   \\
      $t=0.212556$ & & \\
      \hline
      \end{tabular}
    \caption{Comparison of QNM in the dS black hole obtained from the connection formula at 3-instanton in {\bf dictionary 1} with numerical data obtained from the {\it QNMspectral} package.}
    \label{t:CF-6}
\end{table}

\begin{table}
    \centering
    \begin{tabular}{|c|c|c|}\hline
      & 3-instanton & Numerical data \\\hline
      $\Lambda M^2=0.06$ & $\omega_{\rm PS_1}=0.175119-0.0663035i$ & $0.170891-0.063809i$ 
      \\
      $s=l=1$ & $\omega_{\rm PS_2}=0.156801-0.198668i$ & $0.158918-0.193747i$  \\
      $t=0.476876$ & $\omega_{\rm dS_1}=-0.29994i$ & $-0.288i$ \\
      \hline
      $\Lambda M^2=0.08$ & $\omega_{\rm PS_1}=0.13476-0.0505651i$ & $0.133914-0.050199i$ 
      \\
      $s=l=1$ & $\omega_{\rm PS_2}=0.128529-0.15182i$ & $0.128470-0.151125i$ \\
      $t=0.367459$&&\\
      \hline
      $\Lambda M^2=0.1$ & $\omega_{\rm PS_1}=0.0803948-0.0302815i$ & $0.0803459-0.0302718i$  \\
      $s=l=1$ & $\omega_{\rm PS_2}=0.0792948-0.0908679i$ & $0.0792632-0.0908461i$  \\
      $t=0.223449$ & $\omega_{\rm PS_3}=0.0770229-0.15155i$ & $0.077009-0.151528i$  \\
      & $\omega_{\rm dS_1}=-0.440043i$ & $-0.44i$  \\
      \hline
      $\Lambda M^2=0.1$ & $\omega_{\rm PS_1}=0.0256448-0.00303694i$ & $0.0256354-0.00303604i$  \\
      $s=2,l=1$ & $\omega_{\rm PS_2}=0.0256069-0.00911123i$ & $0.0255980-0.00910860i$  \\
      $t=0.223449$ & & \\
      \hline
      \end{tabular}
    \caption{Comparison of QNM in the dS black hole obtained from the connection formula at 3-instanton in {\bf dictionary 2} with numerical data obtained from the {\it QNMspectral} package.}
    \label{t:CF-7}
\end{table}

When the black hole is charged, i.e. for RNdS black hole with
\begin{equation}
    f_{\rm RNdS}(r)=1-\frac{2M}{r}+\frac{Q^2}{r^2}-\frac{r^2}{L^2},
\end{equation}
it has four regular singularities from $f_{\rm RNdS}(r)=0$, and the equation will have five or even six singularities (together with $r=0$ and $r=\infty$) depending on $s$. In this case, more abundant structure of the QNM are found \cite{Cardoso:2017soq,Davey:2022vyx}, and more interesting physics such as the strong cosmic censorship can be explored from the study of QNM. We expect the combination of the quiver gauge theory and the connection formula will tell us more about the QNM (and their analytic properties) in this case, and we leave the details to future works.

\section{Conclusion and discussion}

In this paper, we provide strong supporting evidence that all the families of QNM in the charged C-metric can be reproduced from the connection formula combined with Nekrasov's partition function, which improves the prescription proposed in \cite{Aminov:2020yma}. The radial part of the equation for QNM is shown to be equivalent to the quantized SW equation of SU(2) $N_f=4$ gauge theory, and the angular part is dual to SU(2)$\times$SU(2) superconformal quiver gauge theory. The connection formula gives good numerical approach to all three families of the QNM in the charged C-metric and also allows one to study the QNM analytically as expansion in small parameters. The connection formula between $z=t$ and $z=1$ is shown to be equivalent to the B-cycle quantization condition by assuming all the $\Gamma$-function factors are non-zero, so we conjectured that the missing QNM in the B-cycle quantization approach all come from the poles in the $\Gamma$-function factors in the connection formula. We have shown how to extract out the accelerating modes and the near-extremal modes of the charged C-metric respectively from a $\Gamma$-function factor under the existence of small expansion parameters. The numerical approach through the connection formula between $z=0$ and $z=t$ is also very effective in a rather compensating parameter region, but its physical meaning such as the relation with the B-cycle quantization condition is yet to be uncovered. We also showed that the connection formula can be applied to the dS black hole to recover both the PS modes and the dS modes. We would like to conclude with several proposals for future interesting topics.

\subsection*{Relation with supergroup}

An interesting limit from the gauge theory side is when the matter multiplets
share the pair-wise equal masses
$
    m_1=m_3,\ m_2 =m_4
$. One would expect the gauge theory in this limit to be pathological because the SW curve becomes identical to that of the supergroup gauge theory of SU($2|2$) \cite{Dijkgraaf:2016lym}. The supergroup gauge theory is equivalent to $\hat{A}_1$ affine quiver gauge theory with unphysical gauge coupling $g^2=g_1^2=-g_2^2$, and the energy spectrum is unbounded from the below.  
The dual gravity side is also pathological by analyzing this limit from the map \eqref{eq:map-ai-BHrpm}.
It is straightforward to show the equal pair mass condition will force $r_+=r_-$ whose valid value can only be zero. 
The vanishing horizon size might still allow interesting dynamics, such as Extremal Vanishing horizon (EVH) limit of black holes \cite{Sheikh-Jabbari:2011sar,Chow:2008dp,Balasubramanian:2007bs}.
We leave exploration of these problems in the future work. 

\subsection*{Higher-order differential equations}

It is interesting to ask how to generalize the algorithm of solving eigenvalues of second order differential equation to higher order differential equations.
On the gauge theory side, this corresponds to the SW theory of SU$(N)$ type. 
Quantization of SW curve in these theories will result in differential equation of order $N$ (see e.g. \cite{Nekrasov:2009rc}). The corresponding QNM  problem is then about higher derivative scalar field living in curved black hole spacetimes \cite{Gibbons:2019lmj,Tseytlin:2022flu}.
There are some puzzles about this generalization. 
On the one hand, the QNM are determined by the ingoing boundary condition of the modes at the black hole horizon and boundary condition near the boundary. 
These two conditions together select a regular solution to the second order differential equation and truncate it by appropriate values of eigenenergy.
These two conditions are further related to the quantization condition of the SW curve.
However, how these steps are implemented in differential equations with higher orders, and how these conditions are translated into the quantization condition of the SW theory, should be studied carefully. It has been explored in the literature \cite{Blazquez-Salcedo:2016enn,Blazquez-Salcedo:2017txk,Cardoso:2018ptl,McManus:2019ulj,deRham:2020ejn} that the observation of the second QNM may be used to check the GR and measure the possible extra coupling due to the higher-derivative terms.

\subsection*{Probing Extremal horizon and singularity}

Note the imaginary part of QNM is a diagnostic feature of presence of horizon, thus can be even further applied to study the properties of extremal horizons. 
Other methods can be found in \cite{Cavalcante:2023rdy}. 
One of the typical example is the extremal limit of black holes \cite{Aminov:2020yma}. 
These correspond to the colliding limit of conformal blocks. 
From the gauge theory point of view, one of the chiral multiplet gets infinitely heavy. and decouples itself from the dynamical SCFT. 
This is a doable computation on gauge theory side.
Another possible example is the degenerate horizon (singularity) in hyperscaling violation spacetime \cite{Ogawa:2011bz,Dong:2012se,Shaghoulian:2011aa}.
In the Poincare like coordinate, the hyperscaling violation spacetime metric is 
\begin{equation}\label{eq:hvs}
    ds^2 = -r^{2n} dt^2 + \frac{dr^2}{r^{2m}} +r^2 d\vec{x}^2
\end{equation}
The horizon $r=0$ is the naked singularity and has been shown to be non-singular iff $m=n$ \cite{Copsey:2012gw,Shaghoulian:2011aa}. 
These types of geometries appear in the extremal limit of black brane solutions \cite{Huijse:2011ef,Cremonini:2016bqw,Kiritsis:2016cpm}.
The scalar equation in this background can be shown to be the quantized SW curve of Argyres-Douglas (AD) theory \cite{Kimura:2022yua} of $(A_1,D_{2m+2n-2})$ type, which enables us to study the QNM problem by the corresponding Nekrasov partition function in Argyres-Douglas theory via the AGT conjecture \cite{Nishinaka:2019nuy,Fucito:2023plp}, or through the colliding limit similar to the double scaling limit in the unitary matrix model \cite{Itoyama:2018wbh}. 
As the imaginary part of QNM is due to the presence of horizons, studying this QNM problem by SW algorithm would provide a novel insight into studying the singularity and its resolution in hyperscaling violation (or Lifshitz) spacetimes, including the geometric extension of the horizon in non-singular hyperscaling violation spacetimes \cite{Lei:2013apa}. 
Other spacetimes with singularity that can potentially be probed by this algorithm include Kasner spacetime \cite{1921AmJM} and Janis-Newman-Wincour metric \cite{PhysRevLett.20.878}. Comparing the results obtained from the SW algorithm with the numerical data may also provide some non-trivial consistency check of the AGT conjecture for the AD theories in the NS limit. We will revisit these problems in the future work.

\subsection*{More punctures and quiver gauge theories}

An important point of this article is that the angular part of the charged C-metric can be calculated from the SU(2)$\times$SU(2) quiver gauge theory corresponding to five regular punctures in the class-$\cS$ construction. Only the leading order calculation was presented in this article, and more careful analysis will be reported in our subsequent work together with the application of the related connection formulae to  Kerr-Newman-de Sitter black holes. When the rotation is further included for the black holes, the angular part of the QNM equation will again be challenging to solve \cite{Suzuki:1998vy}, and we wish to borrow the power of the connection formula approach to such problems. The B-cycle quantization condition in quiver gauge theories will be more subtle and losing more families of QNM, but we still expect the connection formula to recover all of them. It is believed to be crucial to develop the connection formula algorithm to study more complicated and general cosmological objects.

\subsection*{QNM from integrable models and wall-crossing}
The quantized Seiberg-Witten curve of ${\cal N}=2$ $SU(2)$ gauge theory with $N_f=0,1,2$ flavours have been studied based on the ordinary differential equation/quantum integrable model (ODE/IM) correspondence \cite{unp-zamo,Fioravanti:2019vxi,Fioravanti:2022bqf}\footnote{See also 
\cite{Ito:2017ypt, Ito:2018eon,Ito:2019jio,Ito:2021boh,Ito:2021sjo} for the case of the Argyres-Douglas theory, which may be related to the QNM of the spacetime \eqref{eq:hvs}.}. The thermodynamic Bethe ansatz (TBA) equations of the corresponding quantum integrable model correspond to the ones of the Gaiotto-Moore-Neitzke determined by the BPS states of the gauge theory \cite{Gaiotto:2010okc,Gaiotto:2014bza, Grassi:2019coc,Grassi:2021wpw}, which provide a powerful method to study the spectral problem of the quantized SW curve. It would be interesting to generalize the TBA equations to the $SU(2)$ gauge theory with $N_f=3,4$ flavours and provide new methods to solve the QNM. Some related attempts can be found in \cite{Caetano:2012ac,Ouyang:2022sje,Fioravanti:2021dce}. It would be very interesting to see the interpretation of the wall-crossing of the BPS states from the viewpoint of the black hole.

\subsection*{Other physical quantities} 

As many physical phenomena, e.g. the tidal deformation and the Hawking radiation, are related to the perturbation theory on the black hole background, one may expect the connection formula to play a more important role in the study of astrophysics and quantum gravity. Indeed the computations of the tidal Love number and the grey-body factor in the Kerr and Kerr-Newman black hole are shown in \cite{Bonelli:2021uvf,Consoli:2022eey}, and it is tempting to apply this trick to many other interesting black hole geometries and explore the astrophysical consequences.

\acknowledgments
We would like to thank Xuefeng Feng, Daniele Gregori, Shuanglin Huang, Jun Nian, Hao Ouyang, Xin Wang, Jingjing Yang, Hongbao Zhang, Hao Zhao and many other people for useful discussions. We would like to thank the organizers of the national conference on Gravitation and Relativistic Astrophysics 2023, the satellite workshop of International Congress of Basic Science, {\it Quantum Gravity and Quantum Field Theory}, and the 4-th national conference for field theory and string theory to allow us to present the results of this work and we benefited a lot from many inspiring discussions during the above workshops. 
Y.L. is supported by a Project Funded by the Priority Academic Program Development of Jiangsu Higher Education Institutions (PAPD) and by National Natural Science Foundation of China No.12305081.
The work of H.S. is supported in part by the Beijing Postdoctoral Research Foundation. K.Z. (Hong Zhang) is supported by a classified fund of Shanghai city. R.Z. is supported by National Natural Science Foundation of China No. 12105198 and the High-level personnel project of Jiangsu Province (JSSCBS20210709).

\appendix

\section{Explicit expressions}\label{a:sol}

\subsection{Conformal blocks at leading orders}

The conformal block can be explicitly evaluated from the three-point structure $\rho(\nu_3,\nu_2,\nu_1|z)$ associated to three primary operators $\nu_{i=1,2,3}$ with conformal weight $h_i$, and the Gram matrix $(G^n_{c,h})^{NM}$ of the Virasoro algebra. More precisely, the three-point correlation function in 2d CFT can be factorized into the holomorphic and anti-holomorphic part as 
\ba
\bra{h_3,\bar{h}_3}V_{h_2,\bar{h}_2}(z,\bar{z})\ket{h_1,\bar{h}_1}=C_{321}C_{\bar{3}\bar{2}\bar{1}}\rho(\nu_3,\nu_2,\nu_1|z)\rho(\bar{\nu}_3,\bar{\nu}_2,\bar{\nu}_1|\bar{z}),
\ea
and the OPE coefficient $C_{321}$ is a theory-dependent structure constant, while $\rho$ is the three-point structure which is determined purely from the representation theory of the Virasoro algebra. In practical, the three-point structure $\rho$ can be computed with the conformal identity, and at the first two levels, 
\begin{align}
\begin{split}
&\rho(\nu_3,\nu_2,L_{-1}\nu_1|z)=(h_1+h_2-h_3)z^{h_3-h_2-h_1-1},\\
&\rho(\nu_3,L_{-1}\nu_2,\nu_1|z)=(h_3-h_2-h_1)z^{h_3-h_2-h_1-1},\\
&\rho(L_{-1}\nu_3,\nu_2,\nu_1|z)=(h_3+h_2-h_1)z^{h_3-h_2-h_1+1},\\
&\rho(L_{-2}\nu_3,\nu_2,\nu_1|z)=(2h_2-h_1+h_3)z^{h_3-h_2-h_1+2},\\
&\rho(L_{-1}L_{-1}\nu_3,\nu_2,\nu_1|z)=(h_3+1+h_2-h_1)(h_3+h_2-h_1)z^{h_3-h_2-h_1+2},\\
&\rho(\nu_3,\nu_2,L_{-1}L_{-1}\nu_1|z)=(h_1+1+h_2-h_3)(h_1+h_2-h_3)z^{h_3-h_2-h_1-2},\\
&\rho(\nu_3,\nu_2,L_{-2}\nu_1|z)=(2h_2+h_1-h_3)z^{h_3-h_2-h_1-2}.
\end{split}
\end{align}

The 4-pt conformal block can be computed via 
\begin{align}
\begin{split}
    &\mathfrak{F}(h_4,h_3,h,h_2,h_1;z)=\cr
    &\sum_{|N|=|M|=n\geq 0}z^{h-h_2-h_1+n}\rho(\nu_4,\nu_3,L_{-N}\nu_{h}|1)(G^n_{c,h})^{NM}\rho(L_{-M}\nu_h,\nu_2,\nu_1|1),\\
    &=:\sum_{n=0}^\infty z^{h-h_2-h_1+n}\mathfrak{F}_n(h_4,h_3,h,h_2,h_1;c),
\end{split}
\end{align}
where $N$ and $M$ are Young diagrams labeling the Verma module of the Virasoro algebra. We have explicitly that 
\begin{align}
    \mathfrak{F}_1=\frac{(h+h_2-h_1)(h+h_3-h_4)}{2h},
\end{align}
and 
\begin{align}
    &2h(c+2ch-10h+16h^2)\mathfrak{F}_2=(4h+8h^2)(2h_3+h-h_4)(2h_2-h_1+h)\cr
&-6h(2h_3+h-h_4)(h+1+h_2-h_1)(h+h_2-h_1)\cr
&-6h(2h_2-h_1+h)(h+1+h_3-h_4)(h+h_3-h_4)\cr
&+(4h+\frac{c}{2})(h+1+h_2-h_1)(h+h_2-h_1)(h+1+h_3-h_4)(h+h_3-h_4).
\end{align}

The 5-pt conformal block is given by 
\begin{align}
    &\mathfrak{F}(h_5,h_4,h_b,h_3,h_a,h_2,h_1;z,t)=\cr
    &\sum_{\substack{|N|=|M|=n\geq 0\\|Y|=|W|=m}}z^{h_a-h_2-h_1+n}t^{h_b-h_3-h_a+m-n}\rho(\nu_5,\nu_4,L_{-W}\nu_{h_b}|1)(G^n_{c,h_b})^{WY}\cr
    &\rho(L_{-Y}\nu_b,\nu_3,L_{-N}\nu_{h_a}|1)(G^n_{c,h_a})^{NM}\rho(L_{-M}\nu_{h_a},\nu_2,\nu_1|1),
\end{align}
and at the sub-leading orders we have 
\begin{align}
    &\mathfrak{F}(h_5,h_4,h_b,h_3,h_a,h_2,h_1;z,t)=z^{h_a-h_2-h_1}t^{h_b-h_3-h_a}\times \cr
    &\lt(1+\frac{(h_a+h_2-h_1)(h_a+h_3-h_b)}{2h_a}\frac{z}{t}+\frac{(h_b+h_3-h_a)(h_4+h_b-h_5)}{2h_b}t+\dots\rt).\cr\label{exp-5pt-conf}
\end{align}

In the semiclassical limit $b\rightarrow 0$, we have at the leading order 
\begin{align}
    &\mathfrak{F}(\Delta_\infty,\Delta_1,\Delta,\Delta_t,\Delta_0;t)=t^{\Delta-\Delta_0-\Delta_t}\exp\lt(b^{-2}F(t)+\widetilde{W}(t)+{\cal O}(b^2)\rt),\\
    &\mathfrak{F}(\Delta_\infty,\Delta_1,\Delta,\Delta_t,\Delta_{0,\pm},\Delta_{2,1},\Delta_0;t)=t^{\Delta-\Delta_{0,\pm}-\Delta_t}z^{\Delta_{0,\pm}-\Delta_0-\Delta_{2,1}}\cr
    &\hskip 6cm \times\exp\lt(b^{-2}F(t)+W(t,z/t)+{\cal O}(b^2)\rt),
\end{align}
where 
\begin{align}
\begin{split}
    &F(t)=\frac{(1-4a^2-4a_1^2+4a_\infty^2)(1-4a^2-4a_t^2+4a_0^2)}{8(1-4a^2)}t+{\cal O}(t^2),\\
    &\widetilde{W}(t)=\frac{1-8a^2+16a^4+16a_0^2(a_1^2-a_\infty^2)-16a_1^2a_t^2+16a_\infty^2a_t^2}{4(1-4a^2)^2}t+{\cal O}(t^2),\\
    &W(t,z/t)=\frac{1-8a^2+16a^4+16a_0^2(a_1^2-a_\infty^2)-16a_1^2a_t^2+16a_\infty^2a_t^2}{4(1-4a^2)^2}t\cr
    &-\frac{1}{2}\frac{a_0(1-4a^2-4a_1^2+4a_\infty^2)}{1-4a^2}t-\frac{1+4a^2-4 a0^2-4 at^2}{4(1+2a_0)}\frac{z}{t}+{\cal O}(t^2,z,z^2/t^2).
\end{split}
\end{align}
From the above explicit expressions, one can confirm that 
\begin{equation}
    W(t,z/t)-\widetilde{W}(t)=-\frac{1}{2}\partial_{a_0}F(t)+{\cal O}(t^2,z,z/t).
\end{equation}

\subsection{Nekrasov's partition function}\label{a:Nekra}

Nekrasov's partition function on the $\Omega$-background is factorized into the perturbative and the instanton part. 
\begin{equation}
    Z=Z^{\rm pert}Z^{\rm inst}.
\end{equation}

The perturbative contribution from the vector multiplet is given by \cite{Nekrasov:2002qd,Losev:2003py}
\begin{align}
    Z^{\rm pert}_{U(N)}(\{\mathfrak{a}_i\})=\Lambda^{\frac{1-N^2}{12}}\exp\lt(\sum_{\alpha:\ {\rm root}}\gamma^{\rm 4d}_{\epsilon_1,\epsilon_2}(\mathfrak{a}_\alpha;\Lambda)\rt),
\end{align}
where
\begin{align}
    \gamma^{\rm 4d}_{\epsilon_1,\epsilon_2}(x;\Lambda)=\lt.\frac{{\rm d}}{{\rm d}s}\lt(\frac{\Lambda^s}{\Gamma(s)}\int_0^\infty {\rm d}t\ t^{s-1}\frac{e^{-tx}}{(e^{\epsilon_1 t}-1)(e^{\epsilon_2 t}-1)}\rt)\rt|_{s=0}.
\end{align}
Taking the Nekrasov-Shatashvili limit with $\epsilon_2=0$ and $\epsilon_1=\hbar$, 
\begin{align}
    \gamma^{\rm 4d}_{\epsilon_1,\epsilon_2}\rightarrow \frac{\gamma_{\rm NS}(x)}{\epsilon_2}=\frac{1}{\epsilon_2}\lt.\frac{{\rm d}}{{\rm d}s}\lt(\frac{\Lambda^s}{\Gamma(s)}\int_0^\infty {\rm d}t\ t^{s-2}\frac{e^{-tx}}{(e^{\hbar t}-1)}\rt)\rt|_{s=0}\cr
    =\frac{1}{\epsilon_2}\lt.\frac{{\rm d}}{{\rm d}s}\lt(\frac{\Lambda^s}{\Gamma(s)}\hbar^{1-s}\int_0^\infty {\rm d}\tilde{t}\ \tilde{t}^{s-2}\frac{e^{-\tilde{t}\frac{x}{\hbar}}}{(e^{\tilde{t}}-1)}\rt)\rt|_{s=0},
\end{align}
then we have
\begin{align}
    \partial_x\gamma_{\rm NS}(x)=\lt.\frac{{\rm d}}{{\rm d}s}\lt(-\frac{\Lambda^s}{\Gamma(s)}\hbar^{-s}\int_0^\infty {\rm d}t\ t^{s-1}\frac{e^{-t\frac{x}{\hbar}}}{(e^{t}-1)}\rt)\rt|_{s=0}\cr
    =\lt.\frac{{\rm d}}{{\rm d}s}\lt(-\lt(\frac{\Lambda}{\hbar}\rt)^s\zeta\lt(s,1+\frac{x}{\hbar}\rt)\rt)\rt|_{s=0}\cr
    =\lt(1-\frac{x}{\hbar}\rt)\log\frac{\Lambda}{\hbar}-\log\Gamma\lt(1+\frac{x}{\hbar}\rt)+\frac{1}{2}\log(2\pi),
\end{align}
where we used 
\begin{equation}
    \zeta(s,x)=\frac{1}{\Gamma(s)}\int_{0}^\infty {\rm d}t\frac{t^{s-1}e^{-xt}}{1-e^{-t}},\quad \zeta(0,x)=\frac{1}{2}-x,
\end{equation}
and the identity
\begin{equation}
    \lt.\frac{\partial}{\partial s}\zeta(s,x)\rt|_{s=0}=\log\Gamma(x)-\frac{1}{2}\log(2\pi).
\end{equation}
Therefore the full perturbative contribution from the vector multiplet of SU(2) gauge theory is given by 
\begin{align}
    \lim_{\epsilon_2\rightarrow 0}\epsilon_2\partial_a F^{\rm pert}_{SU(2)}=-\frac{8a}{\hbar}\log\frac{\Lambda}{\hbar}-2\log\frac{\Gamma\lt(1+\frac{2a}{\hbar}\rt)}{\Gamma\lt(1-\frac{2a}{\hbar}\rt)},
\end{align}
where two Coulomb moduli are chosen to be $\mathfrak{a}_1=-\mathfrak{a}_2=a$. 

Similarly the perturbative contributions from fundamental hypermultiplets are given by 
\begin{equation}
    Z^{\rm pert}_{\rm fund.}(\{\mathfrak{a}_i\},\{m_f\})=\exp\lt(-\sum_{i=1}^N\sum_{f=1}^{N_f}\gamma_{\epsilon_1,\epsilon_2}(\mathfrak{a}_i-m_f)\rt).
\end{equation}
The matter contribution in $\partial_aF^{pert}$ can thus be expressed as 
\begin{equation}
    \lim_{\epsilon_2\rightarrow 0}\epsilon_2\partial_a F^{\rm pert}_{\rm fund.}=\sum_{f=1}^{N_f}\frac{2a}{\hbar}\log\frac{\Lambda}{\hbar}+\log\frac{\Gamma\lt(1+\frac{a-m_f}{\hbar}\rt)}{\Gamma\lt(1-\frac{a+m_f}{\hbar}\rt)}
\end{equation}

The instanton part is composed of the building block usually called the Nekrasov factor, 
\begin{align}
    N_{\lambda\nu}(a,\epsilon_1,\epsilon_2):=\prod_{(i,j)\in\lambda}\lt(a+\epsilon_1(-\nu^t_j+i)+\epsilon_2(\lambda_i-j+1)\rt)\cr
    \times \prod_{(i,j)\in\nu}\lt(a+\epsilon_1(\lambda^t_j-i+1)+\epsilon_2(-\nu_i+j)\rt).
\end{align}
In particular, 
\begin{align}
    &N_{\lambda\emptyset}(a,\epsilon_1,\epsilon_2)=\prod_{(i,j)\in\lambda}\lt(a+i\epsilon_1+\epsilon_2(\lambda_i-j+1)\rt)\rightarrow \prod_{(i,j)\in\lambda}\lt(a+i\hbar\rt),\\
    &N_{\emptyset\lambda}(a,\epsilon_1,\epsilon_2)=\prod_{(i,j)\in\lambda}\lt(a-(i-1)\epsilon_1+\epsilon_2(-\lambda_i+j)\rt)\rightarrow \prod_{(i,j)\in\lambda}(a-(i-1)\hbar),
\end{align}
where the arrow $\rightarrow$ denotes the Nekrasov-Shatashvili limit $\epsilon_1=\hbar$, $\epsilon_2\rightarrow 0$.  
The SU(2) vector multiplet contribution reads 
\begin{equation}
    Z^{\rm vect}_{\mu\nu}(a)=N^{-1}_{\mu\mu}(1,\epsilon_1,\epsilon_2)N^{-1}_{\mu\nu}(2a,\epsilon_1,\epsilon_2)N^{-1}_{\nu\nu}(1,\epsilon_1,\epsilon_2)N^{-1}_{\nu\mu}(-2a,\epsilon_1,\epsilon_2),
\end{equation}
and the fundamental and anti-fundamental matter contributions in SU(2) gauge theories are respectively given by 
\begin{align}
    Z^{\rm fund}_{\mu\nu}(a,\{m_f\}):=\prod_{f=1}^{N_f}N_{\mu\emptyset}(a-m_f,\epsilon_1,\epsilon_2)N_{\nu\emptyset}(-a-m_f,\epsilon_1,\epsilon_2)\cr
    \rightarrow\prod_{f=1}^{N_f}\prod_{(i,j)\in\mu}(a-m_f+i\hbar)\prod_{(i,j)\in\nu}(-a-m_f+i\hbar)\\
   Z^{\rm anti}_{\mu\nu}(a,\{m_f\}):=(-1)^{|\mu|+|\nu|}\prod_{f=1}^{N_{f'}}N_{\emptyset\mu}(-a+m_f-4\epsilon_+,\epsilon_1,\epsilon_2)N_{\emptyset\nu}(a+m_f-4\epsilon_+,\epsilon_1,\epsilon_2)\cr
    \rightarrow (-1)^{|\mu|+|\nu|}\prod_{f=1}^{N_{f'}}\prod_{(i,j)\in\mu}(-a+m_f-2\hbar-(i-1)\hbar)\prod_{(i,j)\in\nu}(a+m_f-2\hbar-(i-1)\hbar)\cr
    =\prod_{f=1}^{N_{f'}}\prod_{(i,j)\in\mu}(a+\hbar-m_f-2\hbar+i\hbar)\prod_{(i,j)\in\nu}(-a+\hbar-m_f+i\hbar),
\end{align}
where $\epsilon_+:=\frac{\epsilon_1+\epsilon_2}{2}$ is a frequently used notation. As can be checked explicitly from the above expressions in the NS limit, the anti-fundamental contribution can in fact be obtained from the fundamental one by substituting $2\epsilon_+-m$ instead. 
\begin{equation}
    Z^{\rm anti}_{\mu\nu}(a,\{m_f\})=Z^{\rm fund}_{\mu\nu}(a,\{2\epsilon_+-m_f\}).
\end{equation}

The full instanton partition function of SU(2) gauge theory with $N_f=4$ flavors is thus given by 
\begin{equation}
    Z^{\rm inst}_{SU(2)\ N_f=4}(a,\{m_f\}_{f=1}^4;\mathfrak{q})=\sum_{\mu,\nu}\mathfrak{q}^{|\mu|+|\nu|}Z^{\rm vect}_{\mu\nu}(a)Z^{\rm fund}_{\mu\nu}(a,\{m_f\}_{f=1}^4),
\end{equation}
where $\Lambda=\mathfrak{q}$ in this case and we put all flavors to be fundamental ones, but as noted before one can always convert any hypermultiplet to an anti-fundamental one by flipping the mass parameter $m\rightarrow 2\epsilon_+-m$.

\subsection{Angular part of charged C-metric}\label{a:angular}

In this section, we briefly describe how to solve the angular equation \eqref{eq:BHODE-theta} purely numerically, through the perturbation approach with respect to $\alpha$, and mapping to a quantized SW differential equation. The concrete form of the ODE is given by 
\begin{equation}
    P\sin\theta\frac{{\rm d}}{{\rm d}\theta}\lt(P\sin\theta\frac{{\rm d}}{{\rm d}\theta}\chi(\theta)\rt)+V_\theta\chi(\theta)=m^2\chi(\theta).\label{eq:angular}
\end{equation}

\paragraph{Numerical results}

It is more convenient to change the variable to $x=\cos\theta$, then 
\begin{equation}
    P(x)=1-2\alpha M x+\alpha^2Q^2x^2,\quad \tilde{P}(x):=(1-x^2)P(x),
\end{equation}
and the differential equation becomes 
\begin{equation}
    \tilde{P}(x)\frac{{\rm d}}{{\rm d}x}\lt(\tilde{P}(x)\frac{{\rm d}}{{\rm d}x}\chi(x)\rt)+(V_\theta(x)-m_0^2(1+2M\alpha+\alpha^2Q^2))\chi(x)=0.
\end{equation}
The boundary conditions for the angular part are imposed as 
\begin{align}
    \chi(x)\sim\begin{cases}
e^{mx_\ast} & x\to 1\\
e^{-mx_{\ast}}& x\to -1
\end{cases},
\end{align}
where ${\rm d}x_\ast=-\frac{{\rm d}x}{\tilde{P}(x)}$. This equation has a very similar form to the radial part, so one can also solve it with the {\it QNMspectral} package. We note that one only inputs the value of $m_0$ without specifying a hidden quantum number $\ell$ that coincides with the angular momentum in the limit $\alpha\rightarrow0$ in the computation. In the limit $\alpha\rightarrow 0$, $P(\theta)\rightarrow 1$, and $V_\theta=\lt(\lambda-\frac{1}{3}\rt)\sin^2\theta$, the eigenfunction is thus nothing but the spherical harmonic function, and we have 
\begin{equation}
    \lambda\rightarrow \ell(\ell+1)+\frac{1}{3}.
\end{equation}
We may thus choose the eigenvalue $\lambda$ closed to $\ell(\ell+1)+\frac{1}{3}$ given $\alpha$ not too large, and one expect that $\ell$ gives the ascending ordering of all allowed eigenvalues of $\lambda$. Some numerical results obtained in this way are listed in Table \ref{t:lambda1} and \ref{t:lambda2}. 

\begin{table}
    \centering
    \begin{tabular}{|c|c|c|c|}
    \hline
       & $\alpha M=0.05$ & $\alpha M=0.1$ & $\alpha M=0.1$ \\
       & $Q/M=0.3$ & $Q/M=0.3$ & $Q/M=0.8$ \\
    \hline
      $m_0=0$ & $\lambda=0.331738(\ell=0)$ & $\lambda=0.326900(\ell=0)$ &  $\lambda=0.328775(\ell=0)$ \\
       & $\lambda=2.324596(\ell=1)$ & $\lambda=2.298118(\ell=1)$ & $\lambda=2.305305(\ell=1)$  \\
    & $\lambda=6.309967(\ell=2)$ & $\lambda=6.239155(\ell=2)$ & $\lambda=6.257812(\ell=2)$ \\
    & $\lambda=12.288080(\ell=3)$ & $\lambda=12.150942(\ell=3)$ & $\lambda=12.186665(\ell=3)$ \\
    \hline
    $m_0=1$ & $\lambda=2.646108(\ell=1)$ & $\lambda=2.984735(\ell=1)$ & $\lambda=3.002191(\ell=1)$\\
    & $\lambda=6.852947(\ell=2)$ & $\lambda=7.417384(\ell=2)$ & $\lambda=7.444401(\lambda=2)$ \\
    \hline
    $m_0=2$ & $\lambda=7.375460(\ell=2)$ & $\lambda=8.498292(\ell=2)$ & $\lambda=8.557553(\ell=2)$\\
    & $\lambda=13.796184(\ell=3)$ & $\lambda=15.390622(\ell=3)$ & $\lambda=15.464319(\ell=3)$\\
    \hline
    \end{tabular}
    \caption{Eigenvalues $\lambda$ obtained from the application of {\it QNMspectral} to the angular part. }
    \label{t:lambda1}
\end{table}

\begin{table}
    \centering
    \begin{tabular}{|c|c|c|c|}
    \hline
       & $\alpha M=0.3$ & $\alpha M=0.5$ & $\alpha M=0.3$ \\
       & $Q/M=0.8$ & $Q/M=0.8$ & $Q/M=0.999$\\
    \hline
    $m_0=0$ & $\lambda=0.290224(\ell=0)$ & $\lambda=0.194511(\ell=0)$ & $\lambda=0.303261(\ell=0)$\\
    & $\lambda=2.069931(\ell=1)$ & $\lambda=1.500135(\ell=1)$ & $\lambda=2.123041(\ell=1)$\\
    & $\lambda=5.623913(\ell=2)$ & $\lambda=4.092418(\ell=2)$ & $\lambda=5.762571(\ell=2)$\\
    & $\lambda=10.955768(\ell=3)$ & $\lambda=7.983622(\ell=3)$ & $\lambda=5.762570(\ell=3)$\\
    \hline
    $m_0=1$ & $\lambda=4.773332(\ell=1)$ & $-$ & $\lambda=4.897025(\ell=1)$ \\
    & $\lambda=10.526705(\ell=2)$ &  & $\lambda=10.692405(\ell=2)$ \\
    \hline
    $m_0=2$ & $\lambda=14.30287(\ell=2)$ & $-$ & $\lambda=14.73839(\ell=2)$\\
    & $\lambda=23.9136(\ell=3)$ & & $\lambda=24.43500(\ell=3)$ \\
    \hline
    \end{tabular}
    \caption{Eigenvalues $\lambda$ obtained from the application of {\it QNMspectral} to the angular part. For some unknown reason, we did not obtain any non-trivial result for $\alpha M=0.5$, $Q/M=0.8$, $m_0=1,2$.}
    \label{t:lambda2}
\end{table}

\paragraph{perturbative approach}

As long as $\alpha$ is small, we expect to solve the angular part as an expansion over $\alpha$ starting from the spherical harmonics. Expanding $\chi(\theta)$ and the eigenvalue $\lambda$ in terms of $\alpha$, 
\begin{equation}
    \chi(\theta)=\chi_0(\theta)+\sum_{i=1}^\infty\chi_i(\theta)\alpha^i,\quad \lambda=\ell(\ell+1)+\frac{1}{3}+\sum_{i=1}^N\lambda_i\alpha^i,
\end{equation}
we have 
\begin{align}
    {\cal L}_0\chi_0(\theta)+\alpha\lt({\cal L}_0\chi_1+2M(\ell(\ell+1)+1)\cos\theta\sin^2\theta\chi_0-4Mm_0^2(1-\cos\theta)\chi_0\rt.\cr
    \lt.+2M\sin^3\theta\frac{{\rm d}}{{\rm d}\theta}\chi_0+\lambda_1\sin^2\theta\chi_0\rt)+{\cal O}(\alpha^2)=0,
\end{align}
where we defined 
\begin{equation}
    {\cal L}_0:=\sin\theta\frac{{\rm d}}{{\rm d}\theta}\sin\theta\frac{{\rm d}}{{\rm d}\theta}+\lt(\ell(\ell+1)\sin^2\theta-m^2_0\rt),
\end{equation}
and $\chi_0$ is a spherical harmonic s.t. ${\cal L}_0\chi_0=0$. Changing the variable to $x=\cos\theta$, we obtain 
\begin{align}
    {\cal L}_0\chi_1+2M(\ell(\ell+1)+1)x(1-x^2)\chi_0-4Mm_0^2(1-x)\chi_0-2M(1-x^2)^2\frac{{\rm d}}{{\rm d}x}\chi_0\cr
    +\lambda_1(1-x^2)\chi_0=0.\label{eq:alpha-1}
\end{align}
For example, when we take $\ell=m_0=2$, 
\begin{equation}
    \chi_0=3(1-x^2),
\end{equation}
the equation reduces to 
\begin{equation}
    {\cal L}_0\chi_1+54Mx(1-x^2)^2-48M(1-x)(1-x^2)+3\lambda_1(1-x^2)^2=0.
\end{equation}
$\chi_1$ is expected to be given by a polynomial with a logarithmic term, 
\begin{equation}
    \chi_1=6(1-x^2)\log(1-x)+\sum_{i}c_ix^i,\label{alpha-ansatz-22}
\end{equation}
as the logarithmic asymptotic behavior can be read off from the $\alpha$-expansion of the boundary condition, 
\begin{equation}
    \chi(x)\sim e^{\pm my},\quad x\rightarrow \pm 1,
\end{equation}
where 
\begin{equation}
    {\rm d}y=\frac{{\rm d}\theta}{P(\theta)\sin\theta}=-\frac{{\rm d}x}{(1-\alpha' x+\alpha^{\prime 2}Q^{\prime 2}x^2)(1-x^2)},  
\end{equation}
and for convenience, we set $\alpha':=2\alpha M$, $Q':=Q/2M$. 
In terms of $x$, the boundary conditions are rewritten to 
\begin{align}
    \chi(x)\sim& (1-x)^{\frac{m_0}{2}}+m_0\alpha'(1-x)^{\frac{m_0}{2}}\log(1-x)\cr
    &+\frac{m_0}{2}\alpha^{\prime 2}(1-x)^{\frac{m_0}{2}}\lt(2\log(1-x)+m_0\log^2(1-x)\rt)+{\cal O}(\alpha^{\prime 3}),\quad x\rightarrow 1,\cr
    \\
    \chi(x)\sim& (1+x)^{\frac{m_0}{2}},\quad x\rightarrow -1.
\end{align}
At the level of $\alpha'$, one can easily solve \eqref{eq:alpha-1} under the ansatz \eqref{alpha-ansatz-22} to obtain 
\begin{align}
\begin{split}
&    \chi_1=6(1-x^2)\log(1-x)-\frac{9}{2}x^3+\frac{9}{2}x,\\
 &   \lambda=6+\frac{1}{3}+10\alpha'+{\cal O}(\alpha^{\prime 2}).
    \end{split}
\end{align}
It is straightforward to further proceed to the second level of perturbation to find 
\begin{equation}
    \chi_2=6(1-x^2)\log(1-x)\lt(1+\frac{3}{2}x+\log(1-x)\rt)+\sum_id_ix^i,
\end{equation}
and 
\begin{equation}
    \lambda=6+\frac{1}{3}+10\alpha'+\frac{171+394Q^{\prime 2}}{42}\alpha^{\prime 2}+{\cal O}(\alpha^{\prime 3}).
\end{equation}

In the case of $\ell=m_0=0$, we obtain 
\begin{equation}
    \lambda=\frac{1}{3}-\frac{1}{6}(1-2Q^{\prime 2})\alpha^{\prime 2}+{\cal O}(\alpha^{\prime 3}).
\end{equation}
For $\alpha'=0.1$, $Q'=0.15$, we reproduced $\lambda\simeq 0.332$, and for $\alpha'=0.2$, $Q'=0.15$, $\lambda\simeq 0.327$ is recovered. 

For $\ell=1$, $m_0=0$, we have 
\begin{equation}
    \lambda=2+\frac{1}{3}-\lt(\frac{9}{10}-\frac{19}{15}Q^{\prime 2}\rt)\alpha^{\prime 2}+{\cal O}(\alpha^{\prime 3}),
\end{equation}
which also reproduces known numerical results in the literature at good accuracy. 

\paragraph{standard Schr\"odinger form}

By converting the angular equation \eqref{eq:angular} to the standard Schr\"odinger equation form \eqref{eq:SW-eq} with the variable $y=\cos\theta$, it turns out to have five regular singularities locating at 
\begin{equation}
    y=-1,\quad y=1,\quad y=\frac{M-\sqrt{M^2-Q^2}}{Q^2\alpha},\quad y=\frac{M+\sqrt{M^2-Q^2}}{Q^2\alpha},\quad y=\infty.
\end{equation}
It is more standard to further put three singularities respectively at $z=0$, $1$ and $\infty$ and the remaining ones at $z=t$ and $z=q$, so one can use for example the variable $z=\frac{t(y+1)}{2}$ with 
\begin{equation}
    t=\frac{2\alpha(M+Q^2\alpha-\sqrt{M^2-Q^2})}{1+2M\alpha+Q^2\alpha^2},
\end{equation}
and 
\begin{align}
    q=\frac{(M+Q^2\alpha-\sqrt{M^2-Q^2})^2}{Q^2(1+2M\alpha+Q^2\alpha^2)}.
\end{align}
Schematically we have 
\begin{align}
    Q_{\rm SW}(z)=\frac{\alpha_5}{z^2}+\frac{\beta_5}{(z-t)^2}+\frac{\gamma_5}{(z-q)^2}+\frac{\delta_5}{(z-1)^2}+\frac{\eta_5}{z(z-1)}\cr
    +\frac{\kappa_5}{z(z-t)}+\frac{\mu_5}{z(z-q)},\label{angular-Q}
\end{align}
where $\alpha_5$, $\beta_5$, $\gamma_5$, $\delta_5$, $\eta_5$, $\kappa_5$ and $\mu_5$ are functions of $\alpha$, $M$ and $Q$. 

It is again deeply related to the 5-pt BPZ equation in 2d CFT. In general, the $n$-pt BPZ equation can be written as 
\begin{equation}
    \lt(b^{-1}\frac{\partial^2}{\partial z^2}+\sum_{i=1}^n\frac{\partial_{t_i}}{z-t_i}+\frac{\Delta_i}{(z-t_i)^2}\rt)\langle V_{2,1}(z)\prod_{i=1}^nV_{i}(t_i)\rangle=0.
\end{equation}
Let us first bring it to the familiar form as \eqref{eq:4pt-BPZ} in the case of $n=5$, by putting say $t_1=0$, $t_4=1$ and $t_5=\infty$. 
\begin{align}
    \lt(b^{-2}\partial_z^2+\frac{\Delta_4}{(z-1)^2}-\frac{\sum_{i=1}^4\Delta_i+\Delta_{2,1}+z\partial_z-\Delta_5+t\partial_t+q\partial_q}{z(z-1)}+\frac{\Delta_2}{(z-t)^2}+\frac{\Delta_3}{(z-q)^2}\rt.\cr
    \lt.+\frac{t\partial_t}{z(z-t)}+\frac{q\partial_q}{z(z-q)}-\frac{\partial_z}{z}+\frac{\Delta_1}{z^2}\rt)\cdot\bra{\Delta_4}V_2(1)V_3(t)V_4(q)\Phi(z)\ket{\Delta_1}=0.\label{eq:5pt-BPZ}
\end{align}
Taking the semiclassical limit $b\rightarrow 0$ gives a Schr\"odinger equation with the potential of the form \eqref{angular-Q}, and the dictionary reads 
\begin{align}
    \Delta_1=\frac{1}{4}-d_0^2,\quad \Delta_2=\frac{1}{4}-d_t^2, \quad \Delta_3=\frac{1}{4}-d_q^2,\quad \Delta_4=\frac{1}{4}-d_1^2,
\end{align}
with $y_\pm:=\frac{1}{r_\mp \alpha}$, 
\begin{align}
\begin{split}
    d_0=\frac{my_+y_-}{2(1+y_+)(1+y_-)},\quad d_t=\frac{my_+y_-}{2(1-y_-)(1-y_+)},\\
    d_q=\frac{my_+y_-}{(1-y_-^2)(y_+-y_-)},\quad d_1=\frac{my_+y_-}{(1-y_+^2)(y_+-y_-)},
    \end{split}
\end{align}
and 
\begin{align}
    \kappa_5=\frac{5y_++y_-(5-3y_+)-7}{6(1-y_+)(1-y_-)}-\frac{y_+y_-}{(1-y_-)(1-y_+)}\lambda+\frac{y_+^2y_-^2(5+y_-(y_+-3)-3y_+)}{2(1-y_+)^3(1-y_-)^3}m^2,\cr
    \mu_5=\frac{1-3y_-^2+2y_+y_-}{6(1-y_-)(y_+-y_-)}-\frac{y_+y_-}{(1-y_-)(y_+-y_-)}\lambda-\frac{2y_+^2y_-^2(1-3y_-^2+2y_+y_-)}{(1-y_-)^3(1+y_-)^2(y_+-y_-)^3},\cr
    \eta_5=\frac{1+2y_+y_--3y_+^2}{6(1-y_+)(y_+-y_-)}+\frac{y_+y_-}{(y_+-y_-)(1-y_+)}\lambda+\frac{2y_+^2y_-^2(1+2y_+y_--3y_+^2)}{(y_+-y_-)^3(1-y_+)^3(1-y_+)^2}m^2.
\end{align}
Combining the Matone relation at the leading order \eqref{Matone-5pt-1} and the connection formula \eqref{5-pt-conn} in the semiclassical limit $b\rightarrow 0$, one can estimate $\lambda$ numerically to obtain its correct order. For example, for $\alpha M=0.3$, $Q/M=0.8$, we have $\lambda_1\simeq 5.6$ and $\lambda_2\simeq 9.7$ for the first two eigenvalues at $m_0=1$ compared with the numerical results $\lambda_1=4.773332$ and  $\lambda_2=10.526705$. More details will be reported in our subsequent paper including precise analysis with higher-instanton corrections and more careful dealing with the connection formula. 

\subsubsection{Matone relations}\label{a:Matone}

The Coulomb branch parameter $\bar{a}$ does not directly appear in the differential equation \eqref{eq:SW-eq} and \eqref{QSW}. To relate it to the parameters in the black hole, we need to use the relation \eqref{u-cond} 
\begin{equation}
    u=\lim_{b\rightarrow 0}b^2t\partial_t\log\mathfrak{F}(\a_4,\a_2,\a,\a_3,\a_1;t),
\end{equation}
appearing in the semiclassical limit of the BPZ equation, which is often referred as the Matone relation \cite{Matone:1995rx}. 
For the corresponding $N_f=4$ theory, $t=\Lambda=\mathfrak{q}$ is just the instanton counting parameter, so 
\begin{align}
    u=&-\bar{a}^2+\frac{-2(\bar{m}_1+\bar{m}_2)+2(\bar{m}_1^2+\bar{m}_2^2)}{4}+(\bar{m}_1+\bar{m}_2)\lt(1-\frac{\bar{m}_3+\bar{m}_4}{2}\rt)\frac{\mathfrak{q}}{1-\mathfrak{q}}\cr
    &+\lim_{b\rightarrow 0}b^2\mathfrak{q}\partial_{\mathfrak{q}}\log  Z^{inst}_{SU(2)\ N_f=4}(a,\{2\epsilon_+-m_1,2\epsilon_+-m_2,m_3,m_4\};\mathfrak{q}).
\end{align}
One can solve $\bar{a}^2$ inversely as an expansion series over $\mathfrak{q}$, 
\begin{align}
    \bar{a}=&\frac{1}{2}\sqrt{-1+4a_0^2+4a_t^2-4u}\cr
    &-\frac{(-1+2a_0^2+2a_1^2-2a_\infty^2+2a_t^2-2u)(-1+4a_t^2-2u)}{4(-1+2a_0^2+2a_t^2-2u)\sqrt{-1+4a_0^2+4a_t^2-4u}}\mathfrak{q}+{\cal O}(\mathfrak{q}^2).\label{Matone-u-a}
\end{align}

Similar relations can be promoted to the case of 5-pt BPZ equation \eqref{eq:5pt-BPZ} in the semiclassical limit $b\rightarrow 0$. For the identifications 
\begin{align}
\begin{split}
    \kappa_5&=\lim_{b\rightarrow 0}b^2t\partial_t\log\mathfrak{F}(\Delta_\infty,\Delta_1,\Delta_b,\Delta_q,\Delta_a,\Delta_t,\Delta_0;t,q),\\
    \mu_5&=\lim_{b\rightarrow 0}b^2q\partial_q\log\mathfrak{F}(\Delta_\infty,\Delta_1,\Delta_b,\Delta_q,\Delta_a,\Delta_t,\Delta_0;t,q),
    \end{split}
\end{align}
one can also solve them for $\bar{a}:=\sqrt{\frac{1}{4}-\Delta_a}$ and $\bar{b}:=\sqrt{\frac{1}{4}-\Delta_b}$ with the explicit expression of the 5-pt conformal block, \eqref{exp-5pt-conf} to obtain at the leading orders, 
\begin{align}
    \bar{a}=\sqrt{\frac{1}{4}-\kappa_5-\Delta_t-\Delta_0}+\frac{\mu_5(\kappa_5+2\Delta_t)}{2(\Delta_0+\Delta_t+\kappa_5)\sqrt{1-4\Delta_0-4\Delta_t-4\kappa_5}}\frac{t}{q}+\cdots,\label{Matone-5pt-1}\\
    \bar{b}=\sqrt{\frac{1}{4}-\mu_5-\kappa_5-\Delta_q-\Delta_t-\Delta_0}+\cdots\label{Matone-5pt-2}
\end{align}

\section{A sketch of derivation of a connection formula}\label{a:conn-form}

In this Appendix, we briefly describe the derivation of a connection formula that is used in the calculation of the QNM both in the charged C-metric and the dS black hole with dictionary 2. We refer to \cite{Bonelli:2022ten} for more details on the derivations and more connection formulae.  
We will use the following identities in the derivation, which are rather straightforward to be checked with the explicit expressions of $\cM$, \eqref{M-def}, and OPE coefficients $C^\alpha_{\beta\gamma}$ and $C_{\alpha\beta\gamma}$,
\begin{align}
    \sum_{\theta=\pm}C^{\alpha_{0\theta}}_{\alpha_{2,1}\alpha_0} C_{\alpha_\infty\alpha_1\alpha_{0\theta}} \cM_{\theta\theta'}(b\alpha_0,b\alpha_1;b\alpha_\infty)\cM^\ast_{\theta\theta'}(b\alpha_0,b\alpha_1;b\alpha_\infty)\cr
    =C^{\alpha_{1\theta'}}_{\alpha_{2,1}\alpha_1}C_{\alpha_\infty\alpha_{1\theta'}\alpha_0},\\
    \sum_{\theta,\theta'=\pm }C^{\alpha_{0\theta}}_{\alpha_{2,1}\alpha_0} C_{\alpha_\infty\alpha_1\alpha_{0\theta}}  \cM_{\theta\theta'}(b\alpha_0,b\alpha_1;b\alpha_\infty)\cM^\ast_{\theta(-\theta')}(b\alpha_0,b\alpha_1;b\alpha_\infty)=0,
\end{align}
where we denoted $\alpha_{i\theta}=\alpha_i-\theta\frac{b}{2}$, 
\begin{align}
C_{\alpha\beta\gamma}=\frac{\Upsilon'_b(0)\Upsilon_b(Q+2\alpha)\Upsilon_b(Q+2\beta)\Upsilon_b(Q+2\gamma)}{\Upsilon_b(Q+\alpha+\beta+\gamma)\Upsilon_b(Q+\alpha+\beta-\gamma)\Upsilon_b(Q+\alpha-\beta+\gamma)\Upsilon_b(Q-\alpha+\beta+\gamma)},
\end{align}
and 
\begin{align}
G_\alpha=\frac{\Upsilon_b(2\alpha+Q)}{\Upsilon_b(2\alpha)},\quad C^\alpha_{\beta\gamma}=G_{\alpha}^{-1}C_{\alpha\beta\gamma}.
\end{align}
The above identities are obtained from the expansion of the following conformal blocks respectively around $z\sim 0$ and $z\sim 1$, 
\begin{align}
     \begin{tikzpicture}
        \draw[ultra thick] (-1,1)--(0,0);
        \draw[ultra thick,dashed] (-1,-1)--(0,0);
        \draw[ultra thick] (2,0)--(0,0);
        \draw[ultra thick] (2,0)--(3,1);
        \draw[ultra thick] (2,0)--(3,-1);
        \node at (-1,1) [left,circle,draw=blue!50,fill=blue!30] {$\Delta_0^{(0)}$};
        \node at (-1,-1) [left,circle,draw=red!50,fill=red!30] {$\Delta_{2,1}^{(z)}$};
        \node at (3,1) [right,circle,draw=blue!50,fill=blue!30] {$\Delta_1^{(1)}$};
        \node at (3,-1) [right,circle,draw=blue!50,fill=blue!30] {$\Delta_\infty^{(\infty)}$};
        \node at (1,0) [above] {$\Delta_{0\theta}$};
        \node at (5,0) {$=\sum_{\theta'}\cM_{\theta\theta'}$};
        \draw[ultra thick] (7,1)--(8,0);
        \draw[ultra thick,dashed] (7,-1)--(8,0);
        \draw[ultra thick] (10,0)--(8,0);
        \draw[ultra thick] (10,0)--(11,1);
        \draw[ultra thick] (10,0)--(11,-1);
        \node at (7,1) [left,circle,draw=blue!50,fill=blue!30] {$\Delta_1^{(1)}$};
        \node at (7,-1) [left,circle,draw=red!50,fill=red!30] {$\Delta_{2,1}^{(z)}$};
        \node at (11,1) [right,circle,draw=blue!50,fill=blue!30] {$\Delta_0^{(0)}$};
        \node at (11,-1) [right,circle,draw=blue!50,fill=blue!30] {$\Delta_\infty^{(\infty)}$};
        \node at (9,0) [above] {$\Delta_{1\theta'}$};
    \end{tikzpicture}
\end{align}
and since the 3-pt conformal block with degenerate insertion is proportional to the hypergeometric function $_2F_1$, the connection formula presented here is equivalent to the one given in \eqref{CF-hyperg}.

Let us first consider the case that relates the 5-pt degenerate conformal block expanded around $z\sim 0$ to that around $z\sim t$. To do so, one may perform a coordinate transformation, $z\rightarrow \frac{z-t}{1-t}$, and the conformal blocks are related by
\begin{align}
    \begin{tikzpicture}
        \draw[ultra thick] (-1,1)--(0,0);
        \draw[ultra thick,dashed] (-1,-1)--(0,0);
        \draw[ultra thick] (0,-1)--(1,0);
        \draw[ultra thick] (2,0)--(0,0);
        \draw[ultra thick] (2,0)--(3,1);
        \draw[ultra thick] (2,0)--(3,-1);
        \node at (-1,1) [left,circle,draw=blue!50,fill=blue!30] {$\Delta_0^{(0)}$};
        \node at (-1,-1) [left,,circle,draw=red!50,fill=red!30] {$\Delta_{2,1}^{(z)}$};
        \node at (0,-1) [below,circle,draw=blue!50,fill=blue!30] {$\Delta_{t}^{(t)}$};
        \node at (3,1) [right,circle,draw=blue!50,fill=blue!30] {$\Delta_1^{(1)}$};
        \node at (3,-1) [right,circle,draw=blue!50,fill=blue!30] {$\Delta_\infty^{(\infty)}$};
        \node at (1.5,0) [above] {$\Delta$};
        \node at (0.5,0) [above] {$\Delta_{0,\theta}$};
        \node at (5,0) {$=g(t)\sum_{\theta'}\cA_{\theta\theta'}$};
        \draw[ultra thick] (7,1)--(8,0);
        \draw[ultra thick,dashed] (7,-1)--(8,0);
        \draw[ultra thick] (8,-1)--(9,0);
        \draw[ultra thick] (10,0)--(8,0);
        \draw[ultra thick] (10,0)--(11,1);
        \draw[ultra thick] (10,0)--(11,-1);
        \node at (7,1) [left,circle,draw=blue!50,fill=blue!30] {$\Delta_t^{(0)}$};
        \node at (7,-1.2) [left,circle,draw=red!50,fill=red!30] {$\Delta_{2,1}^{(\frac{z-t}{1-t})}$};
        \node at (8,-1) [below,circle,draw=blue!50,fill=blue!30] {$\Delta_{0}^{(\frac{t}{t-1})}$};
        \node at (11,1) [right,circle,draw=blue!50,fill=blue!30] {$\Delta_1^{(1)}$};
        \node at (11,-1) [right,circle,draw=blue!50,fill=blue!30] {$\Delta_\infty^{(\infty)}$};
        \node at (9.5,0) [above] {$\Delta$};
        \node at (8.5,0) [above] {$\Delta_{t\theta'}$};
    \end{tikzpicture}
\end{align}
where $g(t)=(t-1)^{\Delta_\infty-\Delta_1-\Delta_t-\Delta_{2,1}-\Delta_0}$ is the factor appears in the coordinate transformation of primary operators. 
The equations satisfied by the connection coefficients $\cA_{\theta\theta'}$ are 
\begin{align}
    \sum_{\theta}C^{\alpha_{0\theta}}_{\alpha_{2,1}\alpha_0}C^{\alpha}_{\alpha_t\alpha_{0\theta}}C_{\alpha_\infty\alpha_1\alpha}\cA_{\theta\theta'}\cA^\ast_{\theta\theta'}=C^{\alpha}_{\alpha_{0}\alpha_{t\theta'}}C^{\alpha_{t\theta'}}_{\alpha_{2,1}\alpha_t}C_{\alpha_\infty\alpha_1\alpha}\\
    \sum_{\theta,\theta'}C^{\alpha_{0\theta}}_{\alpha_{2,1}\alpha_0}C^{\alpha}_{\alpha_t\alpha_{0\theta}}C_{\alpha_\infty\alpha_1\alpha}\cA_{\theta\theta'}\cA^\ast_{\theta(-\theta')}=0.
\end{align}
One soon finds a similar solution to the 3-pt case, 
\begin{equation}
    \cA'_{\theta\theta'}=\cM_{\theta\theta'}(b\alpha_0,b\alpha_t,b\alpha),
\end{equation}
satisfying the above equations, but there is a phase d.o.f. unfixed in the equation $\cA_{\theta\theta'}\rightarrow e^{i\phi(\theta,\theta')}\cA_{\theta\theta'}$. It can be fixed by comparing the leading-power terms in the expansion of the conformal blocks, \eqref{leading-confb}. We then finally obtained 
\begin{equation}
    \cA_{\theta\theta'}=e^{i\pi(\Delta-\Delta_0-\Delta_{2,1}-\Delta_t)}\cM_{\theta\theta'}(b\alpha_0,b\alpha_t,b\alpha),
\end{equation}
where we used 
\begin{equation}
    \Delta_{t,\theta}-\Delta_{t}-\Delta_{2,1}=\frac{bQ}{2}+\theta b\alpha_t.
\end{equation}
One can then adopt the regularization defined in \eqref{semi-reg} on both sides to take the semiclassical limit to have 
\begin{equation}
    \cF_\theta(z;t)=\sum_{\theta'}(t-1)^{\frac{1}{2}}\cM_{\theta\theta'}(\bar{a}_0,\bar{a}_t,\bar{a})\cF_{\theta'}\lt(\frac{t-z}{t-1};\frac{t}{t-1}\rt).
\end{equation}
The prefactor comes from the coordinate transformation and the phase of the degenerate field with conformal weight $\Delta_{2,1}\rightarrow -\frac{1}{2}$ in the semiclassical limit $b\rightarrow 0$. 

The generalization to the connection formula for the conformal block with one more primary operator is straightforward, as only the ``local'' structure changes in the conformal block. More explicitly, we have 
\begin{align}
    \begin{tikzpicture}
        \draw[ultra thick] (-1,1)--(0,0);
        \draw[ultra thick,dashed] (-1,-1)--(0,0);
        \draw[ultra thick] (0,-1)--(1,0);
        \draw[ultra thick] (2,0)--(0,0);
        \draw[ultra thick] (2,0)--(3,1);
        \draw[ultra thick] (2,0)--(3,-1);
        \draw[ultra thick] (2.5,-0.5)--(3.5,0.5);
        \node at (-1,1) [left] {$\Delta_0^{(0)}$};
        \node at (-1,-1) [left] {$\Delta_{2,1}^{(z)}$};
        \node at (0,-1) [below] {$\Delta_{t}^{(t)}$};
        \node at (3,1) [right] {$\Delta_q^{(q)}$};
         \node at (3.5,0.5) [right] {$\Delta_1^{(1)}$};
        \node at (3,-1) [right] {$\Delta_\infty^{(\infty)}$};
        \node at (1.5,0) [above] {$\Delta_a$};
        \node at (2.5,-0.5) [left] {$\Delta_b$};
        \node at (0.5,0) [above] {$\Delta_{0,\theta}$};
        \node at (5,0) {$=\tilde{g}(t)\sum_{\theta'}\cA_{\theta\theta'}$};
        \draw[ultra thick] (7,1)--(8,0);
        \draw[ultra thick,dashed] (7,-1)--(8,0);
        \draw[ultra thick] (8,-1)--(9,0);
        \draw[ultra thick] (10,0)--(8,0);
        \draw[ultra thick] (10,0)--(11,1);
        \draw[ultra thick] (10,0)--(11,-1);
        \draw[ultra thick] (10.5,-0.5)--(11.5,0.5);
        \node at (7,1) [left] {$\Delta_t^{(0)}$};
        \node at (7,-1) [left] {$\Delta_{2,1}^{(\frac{z-t}{1-t})}$};
        \node at (8,-1) [below] {$\Delta_{0}^{(\frac{t}{t-1})}$};
        \node at (11,1) [right] {$\Delta_q^{(q')}$};
        \node at (11,-1) [right] {$\Delta_\infty^{(\infty)}$};
        \node at (9.5,0) [above] {$\Delta_a$};
        \node at (8.5,0) [above] {$\Delta_{t\theta'}$};
        \node at (11.5,0.5) [right] {$\Delta_1^{(1)}$};
    \node at (10.5,-0.5) [left] {$\Delta_b$};
    \end{tikzpicture}
    \label{5-pt-conn}
\end{align}
where $q'=\frac{q-t}{1-t}$ and $\tilde{g}(t)=(t-1)^{\Delta_\infty-\Delta_1-\Delta_q-\Delta_t-\Delta_{2,1}-\Delta_0}$.

\section{Heun equation}\label{a:Heun}

In this Appendix, we summarize the asymptotic behaviors of the solutions to the Heun equation. By comparing the asymptotic behaviors of $\psi$ and $\tilde{\psi}$, solutions to different forms of the Heun equation, one can find out the correct connection formula to solve with. 

The Heun equation \eqref{eq-Heun} reads 
\begin{equation}
    \lt(\frac{d^{2}}{dz^{2}}+\lt(\frac{\tilde{\gamma}}{z}+\frac{\tilde{\delta}}{z-1}+\frac{\tilde{\epsilon}}{z-t}\rt)\frac{d}{dz}+\frac{\tilde{\alpha}\tilde{\beta} z-\tilde{q}}{z(z-1)(z-t)}\rt)\tilde{\psi}(z)=0,
\end{equation}
with the constraint $\tilde{\alpha}+\tilde{\beta}=\tilde{\gamma}+\tilde{\delta}+\tilde{\epsilon}-1$.
This Heun equation is mapped to the Schr\"odinger form \eqref{eq:SW-eq} via the following dictionary, 
\begin{equation}
    \begin{aligned}
    \tilde{\psi}(z)&=(1-z)^{-\tilde{\delta}/2}z^{-\tilde{\gamma}/2}(t-z)^{-\tilde{\epsilon}/2}\psi(z),\\
    a_{0}&=\frac{1-\tilde{\gamma}}{2},\quad a_{1}=\frac{1-\tilde{\delta}}{2},\quad
    a_{t}=\frac{1-\tilde{\epsilon}}{2},\quad a_{\infty}=\frac{\tilde{\alpha}-\tilde{\beta}}{2},\\
    u&=\frac{\tilde{\gamma}\tilde{\epsilon}-2\tilde{q}+2\tilde{\alpha}\tilde{\beta} t-t\tilde{\epsilon}(\tilde{\gamma}+\tilde{\delta})}{2(t-1)},
    \end{aligned}
    \label{dic:Heun-to-SW}
\end{equation}
with the d.o.f. to flip the signs of $a_i$.
We can also express the Heun equation data in terms of $a_i$'s and $u$ as 
\begin{equation}
    \begin{aligned}
    &\tilde{\alpha}=1-a_{t}+a_{\infty}-a_{0}-a_{1},\quad\tilde{\beta}=1-a_{t}-a_{\infty}-a_{0}-a_{1},\\
    &\tilde{\gamma}=1-2a_{0},\quad\tilde{\delta}=1-2a_{1},\quad\tilde{\epsilon}=1-2a_{t},\\
    &\tilde{q}=\frac{1}{2}+u-a_{t}+a_{0}\left(2a_{t}-1\right)+t\big(a_{t}^{2}-a_{\infty}^{2}+a_{0}^{2}+(2a_{1}-1)a_{0}+(a_{1}-1)a_{1}-u\big).
    \end{aligned}
\end{equation}

\subsubsection{Solutions around $z\sim 0$}
Around $z=0$, two independent solutions can be expressed as 
\begin{equation}
    \begin{aligned}
    \tilde{\psi}_{-}^{(0)}=&\text{HeunG}[t,\tilde{q},\tilde{\alpha},\tilde{\beta},\tilde{\gamma},\tilde{\delta},z]\\
    \tilde{\psi}_{+}^{(0)}=&z^{1-\tilde{\gamma}}\text{HeunG}[t,\tilde{q}-(\tilde{\gamma}-1)(\tilde{\delta} t+\tilde{\epsilon}),\tilde{\alpha}- \tilde{\gamma}+1, \tilde{\beta}-\tilde{\gamma}+1,2-\tilde{\gamma}, \tilde{\delta},z],
    \end{aligned}
\end{equation}
and are related by
$
    \Big(\tilde{\psi}_{+}^{(0)}(z)\Big)_{a_{0}\to-a_{0}}=z^{\tilde{\gamma}-1}\tilde{\psi}_{-}^{(0)}(z)
$. Around $z\sim 0$, we find
\begin{equation}
    \begin{aligned}
    \tilde{\psi}_{-}^{(0)}(z)&\sim 1+\frac{\tilde{q}}{t\tilde{\gamma}}z+{\cal O}(z^{2})\\
    \tilde{\psi}_{+}^{(0)}(z)&\sim z^{1-\tilde{\gamma}}+{\cal O}(z^{2-\tilde{\gamma}}),\qquad z\sim 0.
    \end{aligned}
\end{equation}
The HeunG function is analytic in the circle $|z|<R$, where $R$ is the distance between $0$ and the nearest singular point.

These solutions are related to the conformal block via
\begin{equation}
    \begin{aligned}
    \tilde{\psi}_{-}^{(0)}(z)&=z^{-\tilde{\gamma}/2}(1-z)^{-\tilde{\delta}/2}(t-z)^{-\tilde{\epsilon}/2}t^{\frac{1}{2}-a_{t}-a_{0}}e^{-\frac{1}{2}\partial_{a_{0}}F}{\cal F}\bigg(\begin{array}{c}
a_{1}\\
a_{\infty}
\end{array}\begin{array}{c}
a\end{array}\begin{array}{c}
a_{t}\\
\\
\end{array}\begin{array}{c}
a_{0-}\end{array}\begin{array}{c}
a_{2,1}\\
a_{0}
\end{array};t,\frac{z}{t}\bigg)\\
\tilde{\psi}_{+}^{(0)}(z)&=z^{-\tilde{\gamma}/2}(1-z)^{-\tilde{\delta}/2}(t-z)^{-\tilde{\epsilon}/2}t^{\frac{1}{2}-a_{t}+a_{0}}e^{\frac{1}{2}\partial_{a_{0}}F}{\cal F}\bigg(\begin{array}{c}
a_{1}\\
a_{\infty}
\end{array}\begin{array}{c}
a\end{array}\begin{array}{c}
a_{t}\\
\\
\end{array}\begin{array}{c}
a_{0+}\end{array}\begin{array}{c}
a_{2,1}\\
a_{0}
\end{array};t,\frac{z}{t}\bigg),
    \end{aligned}
\end{equation}
or in terms of $\psi_\pm$, 
\begin{equation}
    \begin{aligned}
    \psi_{-}^{(0)}(z)&=t^{\frac{1}{2}-a_{t}-a_{0}}e^{-\frac{1}{2}\partial_{a_{0}}F}{\cal F}\bigg(\begin{array}{c}
a_{1}\\
a_{\infty}
\end{array}\begin{array}{c}
a\end{array}\begin{array}{c}
a_{t}\\
\\
\end{array}\begin{array}{c}
a_{0-}\end{array}\begin{array}{c}
a_{2,1}\\
a_{0}
\end{array};t,\frac{z}{t}\bigg)\\\psi_{+}^{(0)}(z)&=t^{\frac{1}{2}-a_{t}+a_{0}}e^{\frac{1}{2}\partial_{a_{0}}F}{\cal F}\bigg(\begin{array}{c}
a_{1}\\
a_{\infty}
\end{array}\begin{array}{c}
a\end{array}\begin{array}{c}
a_{t}\\
\\
\end{array}\begin{array}{c}
a_{0+}\end{array}\begin{array}{c}
a_{2,1}\\
a_{0}
\end{array};t,\frac{z}{t}\bigg).
    \end{aligned}
\end{equation}
The asymptotic behaviors around $z\sim 0$ of $\psi^{(0)}_\pm$ thus read 
\begin{equation}
    \begin{aligned}
    \psi_{-}^{(0)}(z)&\sim t^{\tilde{\epsilon}/2}z^{\tilde{\gamma}/2}\big(1+{\cal O}(z)\big)=t^{\frac{1}{2}-a_{t}}z^{\frac{1}{2}-a_{0}}+\cdots\\
    \psi_{+}^{(0)}(z)&\sim t^{\tilde{\epsilon}/2}z^{1-\tilde{\gamma}/2}\big(1+{\cal O}(z^{2-\gamma/2})\big)=t^{\frac{1}{2}-a_{t}}z^{\frac{1}{2}+a_{0}}+\cdots.
    \end{aligned}
\end{equation}

\subsubsection{Solution around $z\sim 1$}
Two independent solutions around $z\sim 1$ are
\begin{equation}
    \begin{aligned}
    \tilde{\psi}_{-}^{(1)}&=\left(\frac{z-t}{1-t}\right)^{-\tilde{\alpha}}\text{HeunG}\left[t,\tilde{\alpha}(\tilde{\delta}-\tilde{\beta})+\tilde{q},\tilde{\alpha},-\tilde{\beta}+\tilde{\gamma}+\tilde{\delta},\tilde{\delta},\tilde{\gamma},\frac{t(1-z)}{t-z}\right]\\
    \tilde{\psi}_{+}^{(1)}&=(1-z)^{1-\tilde{\delta}}\left(\frac{z-t}{1-t}\right)^{-\tilde{\alpha}+\tilde{\delta}-1}\\
    &\quad\text{HeunG}\left[t,\tilde{q}-\tilde{\alpha}(\tilde{\delta}-\tilde{\beta})-(\tilde{\delta}-1)(\tilde{\alpha}-\tilde{\beta}+\tilde{\gamma} t+1),\tilde{\alpha}-\tilde{\delta}+1,\tilde{\gamma}-\tilde{\beta}+1,2-\tilde{\delta},\tilde{\gamma},\frac{t(1-z)}{t-z}\right],
    \end{aligned}
\end{equation}
whose asymptotic behavior are given by
\begin{equation}
    \begin{aligned}
    \tilde{\psi}_{-}^{(1)}&\sim 1+\frac{(\tilde{\alpha}\tilde{\beta}-\tilde{q})}{\tilde{\delta}(t-1)}(z-1)+{\cal O}\big((z-1)^{2}\big)\\
    \tilde{\psi}_{+}^{(1)}&\sim(1-z)^{1-\tilde{\delta}}+{\cal O}\big((1-z)^{2-\tilde{\delta}}\big),\qquad z\sim 1 
    \end{aligned}
\end{equation}

The basis $\psi^{(1)}_\pm$ at $z\sim 1$ of the SW Heun equation behaves as
\begin{equation}
    \begin{aligned}
    \psi_{-}^{(1)}&\sim(t-1)^{\tilde{\epsilon}/2}(1-z)^{\tilde{\delta}/2}+\cdots=(t-1)^{\frac{1}{2}-a_{t}}(1-z)^{\frac{1}{2}-a_{1}}+\cdots\\
    \psi_{+}^{(1)}&\sim(t-1)^{\tilde{\epsilon}/2}(1-z)^{1-\tilde{\delta}/2}+\cdots=(t-1)^{\frac{1}{2}-a_{t}}(1-z)^{\frac{1}{2}+a_{1}}+\cdots,
    \end{aligned}
\end{equation}
and are related to the semiclassical conformal block by
\begin{equation}
    \begin{aligned}
    \psi_{-}^{(1)}(z)&=e^{\pm i\pi(a_{1}+a_{t})}(1-t)^{\frac{1}{2}-a_{t}}e^{-\frac{1}{2}\partial_{a_{1}}F}\Big(\big(t(1-t)\big)^{-\frac{1}{2}}(t-z){\cal F}\bigg(\begin{array}{c}
a_{0}\\
a_{t}
\end{array}\begin{array}{c}
a\end{array}\begin{array}{c}
a_{\infty}\\
\\
\end{array}\begin{array}{c}
a_{1-}\end{array}\begin{array}{c}
a_{2,1}\\
a_{1}
\end{array};t,\frac{1-z}{t-z}\bigg)\Big)\\\psi_{+}^{(1)}(z)&=e^{\pm i\pi(-a_{1}+a_{t})}(1-t)^{\frac{1}{2}-a_{t}}e^{\frac{1}{2}\partial_{a_{1}}F}\Big(\big(t(1-t)\big)^{-\frac{1}{2}}(t-z){\cal F}\bigg(\begin{array}{c}
a_{0}\\
a_{t}
\end{array}\begin{array}{c}
a\end{array}\begin{array}{c}
a_{\infty}\\
\\
\end{array}\begin{array}{c}
a_{1+}\end{array}\begin{array}{c}
a_{2,1}\\
a_{1}
\end{array};t,\frac{1-z}{t-z}\bigg).
    \end{aligned}
\end{equation}

\subsubsection{Solution around $z\sim t$}
Around $z=t$, two independent solutions are given by 
\begin{equation}
    \begin{aligned}
    \tilde{\psi}_{-}^{(t)}&=\text{HeunG}\Big(\frac{t}{t-1},\frac{ \tilde{q}- \tilde{\alpha} \tilde{\beta} t}{1-t}, \tilde{\alpha}, \tilde{\beta}, \tilde{\epsilon}, \tilde{\delta,}\frac{z-t}{1-t}\Big)\\
     \tilde{\psi}_{+}^{(t)}&=(t-z)^{1- \tilde{\epsilon}} \\
     &\text{HeunG}\Big(\frac{t}{t-1},\frac{ \tilde{q}- \tilde{\alpha} \tilde{\beta} t}{1-t}-( \tilde{\epsilon}-1)\left( \tilde{\gamma}+\frac{ \tilde{\delta }t}{t-1}\right), \tilde{\alpha}- \tilde{\epsilon}+1, \tilde{\beta}- \tilde{\epsilon}+1,2- \tilde{\epsilon}, \tilde{\delta},\frac{z-t}{1-t}\Big).
    \end{aligned}
\end{equation}
The asymptotic behaviors around $z\sim t$ read 
\begin{equation}
    \begin{aligned}
     \tilde{\psi}_{-}^{(t)}&\sim1+\frac{( \tilde{q}- \tilde{\alpha} \tilde{\beta} t)}{(t-1)t \tilde{\epsilon}}(z-t)+{\cal O}\big((z-t)^{2}\big)\\
     \tilde{\psi}_{+}^{(t)}&\sim(t-z)^{1- \tilde{\epsilon}}+{\cal O}\big((z-t)^{2- \tilde{\epsilon}}\big),
    \end{aligned}
\end{equation}
and in terms of $\psi^{(t)}_\pm$ we have 
\begin{equation}
    \begin{aligned}
   \psi_{-}^{(t)}&\sim(t-z)^{ \tilde{\epsilon}/2}(1-t)^{ \tilde{\delta}/2}t^{ \tilde{\gamma}/2}+\cdots=(t-z)^{\frac{1}{2}-a_{t}}(1-t)^{\frac{1}{2}-a_{1}}t^{\frac{1}{2}-a_{0}}+\cdots\\
   \psi_{+}^{(t)}&\sim(t-z)^{1- \tilde{\epsilon}/2}(1-t)^{ \tilde{\delta}/2}t^{ \tilde{\gamma}/2}+\cdots=(t-z)^{\frac{1}{2}+a_{t}}(1-t)^{\frac{1}{2}-a_{1}}t^{\frac{1}{2}-a_{0}}+\cdots,
    \end{aligned}
\end{equation}
with the following relations to the semiclassical conformal blocks, 
\begin{equation}
    \begin{aligned}
    \psi_{-}^{(t)}(z)&=t^{\frac{1}{2}-a_{0}-a_{t}}(1-t)^{\frac{1}{2}-a_{1}}e^{-\frac{1}{2}\partial_{a_{t}}F}\Big((t-1)^{\frac{1}{2}}{\cal F}\bigg(\begin{array}{c}
a_{1}\\
a_{\infty}
\end{array}\begin{array}{c}
a\end{array}\begin{array}{c}
a_{0}\\
\\
\end{array}\begin{array}{c}
a_{t-}\end{array}\begin{array}{c}
a_{2,1}\\
a_{t}
\end{array};\frac{t}{t-1},\frac{t-z}{t}\bigg)\Big)\\\psi_{+}^{(t)}(z)&=t^{\frac{1}{2}-a_{0}+a_{t}}(1-t)^{\frac{1}{2}-a_{1}}e^{\frac{1}{2}\partial_{a_{t}}F}\Big((t-1)^{\frac{1}{2}}{\cal F}\bigg(\begin{array}{c}
a_{1}\\
a_{\infty}
\end{array}\begin{array}{c}
a\end{array}\begin{array}{c}
a_{0}\\
\\
\end{array}\begin{array}{c}
a_{t+}\end{array}\begin{array}{c}
a_{2,1}\\
a_{t}
\end{array};\frac{t}{t-1},\frac{t-z}{t}\bigg)\Big).
    \end{aligned}
\end{equation}

\subsubsection{Solution around $z\to \infty$}

The solutions around $z\sim \infty$ are
\begin{equation}
    \begin{aligned}
     \tilde{\psi}_{-}^{(\infty)}&=z^{-\tilde{\beta}}\text{HeunG}\left[t,\tilde{q}-\tilde{\alpha}\tilde{\beta}(t+1)+\tilde{\beta}(\tilde{\delta}+t\tilde{\epsilon}),\tilde{\beta},\tilde{\beta}-\tilde{\gamma}+1,-\tilde{\alpha}+\tilde{\beta}+1,\tilde{\alpha}+\tilde{\beta}-\tilde{\gamma}-\tilde{\delta}+1,\frac{t}{z}\right]\\
     \tilde{\psi}_{+}^{(\infty)}&=z^{-\tilde{\alpha}}\text{HeunG}\left[t,\tilde{q}-\tilde{\alpha}\tilde{\beta}(t+1)+\tilde{\alpha}(\tilde{\delta}+t\tilde{\epsilon}),\tilde{\alpha},\tilde{\alpha}-\tilde{\gamma}+1,\tilde{\alpha}-\tilde{\beta}+1,\tilde{\alpha}+\tilde{\beta}-\tilde{\gamma}-\tilde{\delta}+1,\frac{t}{z}\right],
    \end{aligned}
    \end{equation}
whose asymptotic behaviors are
\begin{equation}
    \begin{aligned}
 \tilde{\psi}_{-}^{(\infty)}&\sim z^{-\tilde{\beta}}\left(1+\frac{\tilde{q}-\tilde{\alpha}\tilde{\beta}(t+1)+\tilde{\beta}(\tilde{\delta}+t\tilde{\epsilon})}{z(-\tilde{\alpha}+\tilde{\beta}+1)}+{\cal O}(z^{-2})\right)\\
 \tilde{\psi}_{+}^{(\infty)}&\sim z^{-\tilde{\alpha}}\left(1+\frac{\tilde{q}-\tilde{\alpha}\tilde{\beta}(t+1)+\tilde{\alpha}(\tilde{\delta}+t\tilde{\epsilon})}{z(\tilde{\alpha}-\tilde{\beta}+1)}+{\cal O}(z^{-2})\right),\qquad z\sim \infty.
    \end{aligned}
\end{equation}
The basis at $z\sim \infty$ of the SW Heun equation thus behaves as
\begin{equation}
    \begin{aligned}
    \psi_{-}^{(\infty)}&\sim(-1)^{1-a_{t}-a_{1}}z^{\frac{1}{2}+a_{\infty}}+\cdots\\\psi_{+}^{(\infty)}&\sim(-1)^{1-a_{t}-a_{1}}z^{\frac{1}{2}-a_{\infty}}+\cdots,\quad z\sim\infty.
    \end{aligned}
\end{equation}
and are related to the conformal block by
\begin{equation}
    \begin{aligned}
    \psi_{-}^{(\infty)}(z)&=e^{\pm i\pi(1-a_{1}-a_{t})}e^{-\frac{1}{2}\partial_{a_{\infty}}F}\Big(t^{-\frac{1}{2}}{\cal F}\bigg(\begin{array}{c}
a_{t}\\
a_{0}
\end{array}\begin{array}{c}
a\end{array}\begin{array}{c}
a_{1}\\
\\
\end{array}\begin{array}{c}
a_{\infty-}\end{array}\begin{array}{c}
a_{2,1}\\
a_{\infty}
\end{array};t,\frac{1}{z}\bigg)\Big)\\\psi_{+}^{(\infty)}(z)&=e^{\pm i\pi(1-a_{1}-a_{t})}e^{\frac{1}{2}\partial_{a_{\infty}}F}\Big(t^{-\frac{1}{2}}{\cal F}\bigg(\begin{array}{c}
a_{t}\\
a_{0}
\end{array}\begin{array}{c}
a\end{array}\begin{array}{c}
a_{1}\\
\\
\end{array}\begin{array}{c}
a_{\infty+}\end{array}\begin{array}{c}
a_{2,1}\\
a_{\infty}
\end{array};t,\frac{1}{z}\bigg)\Big).
    \end{aligned}
\end{equation}

\subsection{Connection formulae in terms of Heun functions}

We also list the connection formulae written in terms of $\tilde{\psi}$ and we refer to \cite{Bonelli:2022ten} for their derivations. $\theta$ takes the value $\pm$. 
\begin{align}
    &\tilde{\psi}^{(0)}_\theta(z)=\sum_{\theta'=\pm}(1-t)^{-\frac{\tilde\delta}{2}}t^{(1+\theta)\frac{1-\tilde\gamma}{2}-(1+\theta')\frac{1-\tilde\epsilon}{2}}e^{\frac{\theta}{2}\partial_{a_0}F-\frac{\theta'}{2}\partial_{a_t}F}\cr
    &\times\frac{\Gamma\lt(\theta'(\tilde\epsilon-1)\rt)\Gamma\lt(1+\theta(1-\tilde\gamma)\rt)}{\Gamma\lt(\frac{1}{2}-a+\frac{\theta}{2}(1-\tilde\gamma)-\frac{\theta'}{2}(1-\tilde\epsilon)\rt)\Gamma\lt(\frac{1}{2}+a+\frac{\theta}{2}(1-\tilde\gamma)-\frac{\theta'}{2}(1-\tilde\epsilon)\rt)}\tilde{\psi}^{(t)}_{\theta'}(z),\cr\label{conn-w-0t}\\
    &\tilde{\psi}^{(0)}_\theta(z)=\sum_{\theta'\pm}\lt(\sum_{\sigma=\pm}\frac{e^{\frac{i\pi}{2}(\tilde\delta+\tilde\epsilon)}t^{\frac{\tilde\epsilon+\theta(1-\tilde\gamma)}{2}-\sigma a}e^{\frac{\theta}{2}\partial_{a_0}F-\frac{\theta'}{2}\partial_{a_\infty}F-\frac{\sigma}{2}\partial_{a}F}}{\Gamma\lt(1-\sigma a+\frac{\theta}{2}(1-\tilde\gamma)-\frac{\tilde\epsilon}{2}\rt)\Gamma\lt(-\sigma a+\frac{\theta}{2}(1-\tilde\gamma)+\frac{\tilde\epsilon}{2}\rt)}\rt.\cr
    &\lt.\times\frac{1}{\Gamma\lt(1-\sigma a-\frac{\theta'}{2}(\tilde\alpha-\tilde\beta)-\frac{\tilde\delta}{2}\rt)\Gamma\lt(-\sigma a-\frac{\theta'}{2}(\tilde\alpha-\tilde\beta)-\frac{\tilde\delta}{2}\rt)}\rt)\tilde{\psi}^{(\infty)}_{\theta'}(z),\label{conn-w-0inf}\\
    &\tilde{\psi}^{(1)}_{\theta}(z)=-(1-t)^{\frac{\tilde\epsilon}{2}}\sum_{\theta'=\pm}\frac{\Gamma\lt(1+\theta(1-\tilde\delta)\rt)\Gamma\lt(\theta'(\tilde\beta-\tilde\alpha)\rt)e^{\frac{\theta}{2}\partial_{a_1}F-\frac{\theta'}{2}\partial_{a_\infty}F}}{\prod_{\sigma=\pm}\Gamma\lt(\frac{1}{2}+\sigma a+\frac{\theta}{2}(1-\tilde\delta)+\frac{\theta'}{2}(\tilde\beta-\tilde\alpha)\rt)}\tilde{\psi}^{(\infty)}_{\theta'}(z),\cr\label{conn-w-1inf}
\end{align}
where we still keep the variables $a_i$'s in the derivatives (they can be translated at anytime wtih \eqref{dic:Heun-to-SW}), and $a$ needs to be solved from the Matone relation \eqref{u-cond} (more details can be found in Appendix \ref{a:Matone}). 

\bibliographystyle{JHEP}
\bibliography{BH-SW}

\end{document}